\documentclass[reprint,amsmath,amssymb,aps,superscriptaddress]{revtex4-2}

\usepackage{graphicx}
\usepackage{dcolumn}
\usepackage{bm}
\usepackage[dvipsnames]{xcolor}
\usepackage[colorlinks=true, citecolor=blue, linkcolor=blue]{hyperref}
\colorlet{LightRubineRed}{RubineRed!70!}
\colorlet{Mycolor1}{green!10!orange!90!}
\definecolor{Mycolor2}{HTML}{00F9DE}

\begin{document}

\preprint{APS/123-QED}

\title{Geometric Mechanics of Real-Space Energy Transfer in Turbulence}

\author{Weijie, Zhang}

\author{Hanxun, Yao}\email{hanxun@ustc.edu.cn}
 \affiliation{%
 Department of Modern Mechanics, University of Science and Technology of China, Hefei, China
}%

\date{\today}


\begin{abstract}
How the local energy transfer across scales in turbulence is encoded in the instantaneous deformation of fluid elements remains a fundamental question. We exploit the spherical geometry of the real-space energy cascade at length scale $\ell$. An exact derivation shows that it can be decomposed locally by strain self-amplification and vortex stretching, with a coefficient equal to the specific moment of inertia $2/5(\ell/2)^2$ of a spherical fluid neighborhood. Direct numerical simulations show that this contribution captures the dominant local transfer and defines a statistical boundary separating forward- and inverse-cascade events.  A further longitudinal and transverse decomposition reveals that the longitudinal transfer underlying Kolmogorov's $4/5$
law is governed solely by strain self-amplification, whereas vortex stretching enters through the transverse component. These results establish a purely mechanical connection between turbulent energy cascade and the local rotational and deformational dynamics of fluid elements.
\end{abstract}


\maketitle
The energy cascade is one of the central concepts of turbulence theory. Richardson envisioned kinetic energy being transferred from large eddies to smaller ones until it is dissipated by viscosity \cite{Richardson1922}, and Kolmogorov established the
statistical foundation of this picture \cite{K41}. For homogeneous isotropic turbulence at large Reynolds number, the inertial-range transfer obeys the exact $4/5$ law,
$
\langle[\delta u_L(r)]^3\rangle
=-(4/5)\langle\varepsilon\rangle r,
$
where $\delta u_L(r)$ is the longitudinal velocity with separation $r$, and $\langle\varepsilon\rangle$ is the global averaged energy dissipation rate. Its negative sign reflects the net transfer of kinetic energy toward
smaller scales. Locally, however, the cascade fluctuates strongly and can proceed in either direction. Negative subgrid-scale energy flux,
commonly termed ``backscatter'', has long been observed in
large-eddy simulations and direct numerical simulations
\cite{Piomelli1991,Meneveau2000}. The local fluctuations of forward and inverse transfer have been associated with distinct velocity-gradient configurations \cite{borue1998local,yao2024comparing}.

Identifying the local dynamical mechanisms responsible for such interscale transfer remains a fundamental problem. Vortex stretching (VS) has long been associated with the generation of small scales
\cite{taylor1938production,onsager1949statistical,tennekes1972first}. A complementary description follows from the filtered Navier-Stokes equations. For the subgrid-scale (SGS)
$
\tau_{ij}
=
\widetilde{u_i u_j}
-
\widetilde u_i\widetilde u_j,
$
the local SGS energy flux is
$
\Pi_\ell=-\tau_{ij}\widetilde S_{ij},
$
where
$\widetilde S_{ij}
=(\partial_j\widetilde u_i+\partial_i\widetilde u_j)/2$
and a tilde denotes filtering at scale $\ell$
\cite{germano1992turbulence}. Borue and Orszag
\cite{borue1998local} related $\Pi_\ell$, at leading order, to two fundamental nonlinear velocity-gradient processes, strain self-amplification (SSA) and VS. More recently, Johnson \cite{johnson2020local} derived an exact multiscale decomposition of the SGS flux for Gaussian filter, in which these mechanisms emerge together with contributions from smaller scales.

Yet $\Pi_\ell$ is defined at a fixed filter scale and does not, by itself, arise as a divergence of a flux in real/scale space. A genuine local scale-space flux follows instead from the generalized Kolmogorov-Hill (KH) equation (also called K\'arm\'an--Howarth--Monin--Hill equation \cite{portela2017turbulence}) \cite{hill2002exact}. Integrating this equation over a sphere of diameter $\ell$ in separation space gives the local scale-integrated
energy balance \cite{yao2023entropy,yao2024forward},
$
\hat d k_\ell/dt
=
\Phi_\ell+P_\ell+D_\ell-\epsilon_\ell,
$
where $\hat d /dt$ means material derivatives with convection velocity equal to the two-point average velocity. $k_\ell$ is the kinetic energy associated with scales up to
$\ell$, $P_\ell$ and $D_\ell$ denote pressure-work and viscous
transport, respectively, and $\epsilon_\ell$ is the locally averaged
dissipation rate. The corresponding real-space energy transfer rate is
\begin{equation}
    \Phi_{\ell}(\mathbf{x},t)
    =
    -\frac{3}{4\ell}\frac{1}{S_{\ell}}
    \oint_{S_{\ell}}
    \delta u_i^2\,\delta u_j\hat{n}_j\,dS,
\label{Phi_definition}
\end{equation}
where
$\delta u_i
=u_i(\mathbf{x}+\mathbf r)-u_i(\mathbf{x}-\mathbf r)$,
$r=|\mathbf r|=\ell/2$,
$\hat{\mathbf n}=\mathbf r/r$, and
$S_\ell=4\pi r^2$.
Because $\Phi_\ell$ originates directly from the divergence of the
two-point energy flux in separation space, it provides an
unambiguous local measure of transfer across scale $\ell$
\cite{yao2024comparing} such that $\Phi_\ell>0$ denotes forward transfer and $\Phi_\ell<0$ inverse transfer. Equation~\eqref{Phi_definition} poses a basic dynamical question, how is a genuine local energy flux through real space represented in the instantaneous geometry of the velocity field? Previous work showed that forward and inverse events of $\Phi_\ell$ occupy systematically different regions of velocity-gradient invariants  \cite{yao2024comparing}, while related connections between structure-function transfer, strain, and vorticity dynamics have been established primarily through averaged quantities \cite{carbone2020vortex} so that the inverse events are ignored. A direct local relation between the finite velocity increments constituting $\Phi_\ell$ and the coarse-grained velocity-gradient tensor (VGT), capable of identifying both the mechanism and direction of local energy transfer, has remained unclear.

In this Letter, we establish a direct geometrical and mechanical connection between the real-space energy cascade and coarse-grained VGT dynamics. We show that the finite velocity increments entering the local energy flux admit a spherical harmonic decomposition in which the component resolved by the coarse-grained VGT can be isolated unambiguously. This provides a local
representation of the dominant energy-transfer dynamics in terms of SSA and VS with an exact geometrical coefficient, while the remaining angular information is retained as a residual. Direct numerical simulations show that this VGT-resolved contribution captures the dominant statistical organization of the cascade and provides simple boundaries separating forward and inverse energy transfer.

We start from the geometry of $\Phi_\ell$ itself, which suggests the natural coarse-graining domain. Its spherical surface encloses a volume $V_\ell$ of radius $r$, and we therefore define the
coarse-grained velocity by averaging over the same spherical neighborhood, corresponding to a spherical top-hat filter,
$
    \widetilde u_i(\mathbf{x})
    =
    \frac{1}{V_\ell}
    \int_{V_\ell}
    u_i(\mathbf{x}+\mathbf r)\,d^3\mathbf r .
$
The coarse-graining volume is thus geometrically paired with the
surface integral defining the scale-space flux. Differentiating with
respect to $\mathbf{x}$ and applying Gauss's theorem in
$\mathbf r$-space gives the exact relation (see Supplemental Material)
\begin{equation}
    \ell\widetilde A_{ij}
    =
    \frac{3}{S_\ell}
    \oint_{S_\ell}
    \delta u_i\,\hat n_j\,dS ,
\label{du_VGT_relation}
\end{equation}
which follows purely from this spherical geometry and requires neither
a Taylor expansion \cite{carbone2020vortex} nor an assumption concerning a local H\"older
exponent \cite{borue1998local}.

At fixed $\mathbf{x}$ and $r$, each component
$\delta u_i(\hat{\mathbf n})$ defines an angular field on the sphere and therefore admits a spherical-harmonic decomposition $
    \delta u_i(\hat{\mathbf n})
    =
    \sum_{L=0}^{\infty}
    \delta u_i^{(L)}(\hat{\mathbf n})$, where superscript $L$ denotes the spherical-harmonic degree. Such decompositions have been used previously to separate angular sectors
of turbulent structure functions
\cite{Arad1998,Arad1999a,Arad1999b,Kurien2001}.
The dipolar ($L=1$) subspace is spanned by
$\{\hat n_1,\hat n_2,\hat n_3\}$, so that its projection can be written
as $B_{ij}\hat n_j$, with
$
B_{ij}
=
3S_\ell^{-1}
\oint_{S_\ell}\delta u_i\hat n_j\,dS.
$
Equation~\eqref{du_VGT_relation} therefore gives, exactly,
$
B_{ij}=\ell\widetilde A_{ij}.
$
Hence the finite velocity increment admits the unique decomposition
\begin{equation}
    \delta u_i
    =
    2\widetilde A_{ij}r_j
    +
    q_i(\hat{\mathbf n}),
\label{du_est}
\end{equation}
where the first term is the exact $L=1$ component. Because of $\delta u_i(-\hat{\mathbf n})=-\delta u_i(\hat{\mathbf n})$, all even angular sectors vanish and
$
q_i
=
\sum_{L=3,5,\ldots}\delta u_i^{(L)}
$
contains the remaining higher odd modes. By orthogonality,
$
S_\ell^{-1}
\oint_{S_\ell}q_i\hat n_j\,dS=0.
$
Thus no additional $L=1$ contribution is contained in $q_i$, i.e., the coarse-grained VGT uniquely determines the entire dipolar sector, whereas the higher angular structure is not determined by $\widetilde A_{ij}$ alone. 

We now isolate the contribution of the VGT-resolved dipolar component to the local real-space energy flux. Retaining
$
\delta u_i^{(1)}
=
2\widetilde A_{ij}r_j
$
in Eq.~\eqref{Phi_definition} gives
\begin{equation}
    (\delta u_i^{(1)})^2
    =
    4\|\widetilde{\mathbf A}\mathbf r\|^2,
    \qquad
    \delta u_j^{(1)}\hat n_j
    =
    2\widetilde A_{jk}r_k\hat n_j .
\label{du2_decom_all}
\end{equation}
The real-space energy flux (Spherical integration), followed by decomposition of
$\widetilde A_{ij}$ into its symmetric and antisymmetric parts, then
gives (see Supplemental Material)
\begin{equation}
    \Phi_\ell(\mathbf{x},t)
    =
    \frac{2}{5}(\frac{\ell}{2})^2
    \left(
    3\widetilde R_s
    +
    \frac{1}{4}\widetilde V_s
    \right)
    +
    O_\ell ,
\label{tophat}
\end{equation}
where
$
\widetilde R_s
=
-\widetilde S_{ij}\widetilde S_{jk}\widetilde S_{ki}/3
$
and
$
\widetilde V_s
=
\widetilde\omega_i
\widetilde S_{ij}
\widetilde\omega_j
$
represent SSA and VS at length scale $\ell$, respectively, with
$
\widetilde{\boldsymbol{\omega}}
=
\nabla\times\widetilde{\mathbf u}
$
the coarse-grained vorticity. Here $O_\ell$ contains all contributions involving the higher angular
modes $q_i$. For comparison, Borue and Orszag \cite{borue1998local} introduced H\"older continuous ``Fractal Taylor Expansion'', under which the top-hat filter gives the filtering energy cascade, 
$
    \Pi_{\ell}(\mathbf{x},t)
    \approx
    \frac{3}{11} \left(\frac{\ell}{2}\right)^2
    \left(
    3\widetilde{R}_s
    +
    \frac{1}{4}\widetilde{V}_s
    \right).
$ Hence, although the structure-function and filtering formulations are
defined through different energy budgets, they involve the same
leading-order local velocity-gradient invariants, i.e., 
$
    \Phi_\ell,\Pi_\ell
    \propto
    3\widetilde{R}_s+\frac{1}{4}\widetilde{V}_s,
$ but different prefactors.


We use data from the Johns Hopkins Turbulence Database,
\texttt{isotropic8192} dataset, a direct numerical simulation of forced homogeneous isotropic turbulence on an $8192^3$ grid at $Re_\lambda\approx1250$. Five length scales, $\ell/\eta=15$, $30$, $45$, $60$, and $75$, are considered, spanning the viscous to inertial ranges, where $\eta=(\nu^3/\langle\epsilon\rangle)^{1/4}$ is the Kolmogorov
length scale. The $\widetilde A_{ij}$ is calculated through spherical top-hat filter velocity and fourth-order central difference. The local energy-transfer rate $\Phi_\ell$ is evaluated by discretizing the surface integral of
$\delta u_i^2\delta u_j\hat n_j$ following the procedure described in \cite{yao2025analysis}. Approximately uniformly distributed pairs of opposite points are used on each sphere, with 300 surface samples
at $\ell=15\eta$, 500 at $\ell/\eta=30$, $45$, and $60$, and 2000 at $\ell=75\eta$. Statistics are collected from $2\times10^6$ spheres whose centers are uniformly and randomly distributed throughout the computational domain. Convergence with respect to the
number of surface points and statistical sampling was verified to ensure the robustness of the reported results.

\begin{figure}[htbp]
    \centering
    \includegraphics[width=0.48\textwidth]{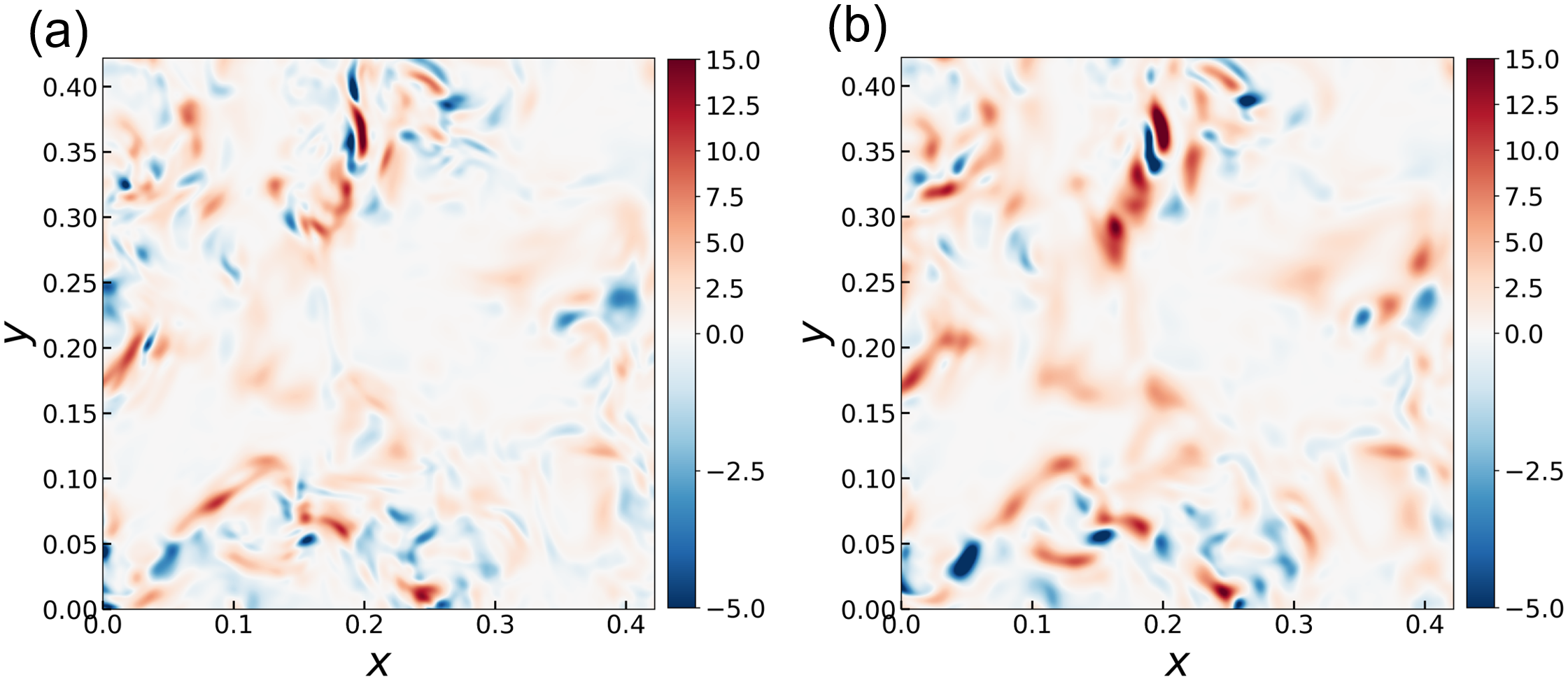}
    \caption{Instantaneous spatial distributions at
    $\ell=45\eta$ of panel (a) the local scale-space energy flux
    $\Phi_\ell$ and panel (b) its leading VGT approximation
    $(2/5)X_\ell$.}
    \label{fig:Phi_SSAVS}
\end{figure}

\begin{figure}[htbp]
    \centering
    \includegraphics[width=0.35\textwidth]{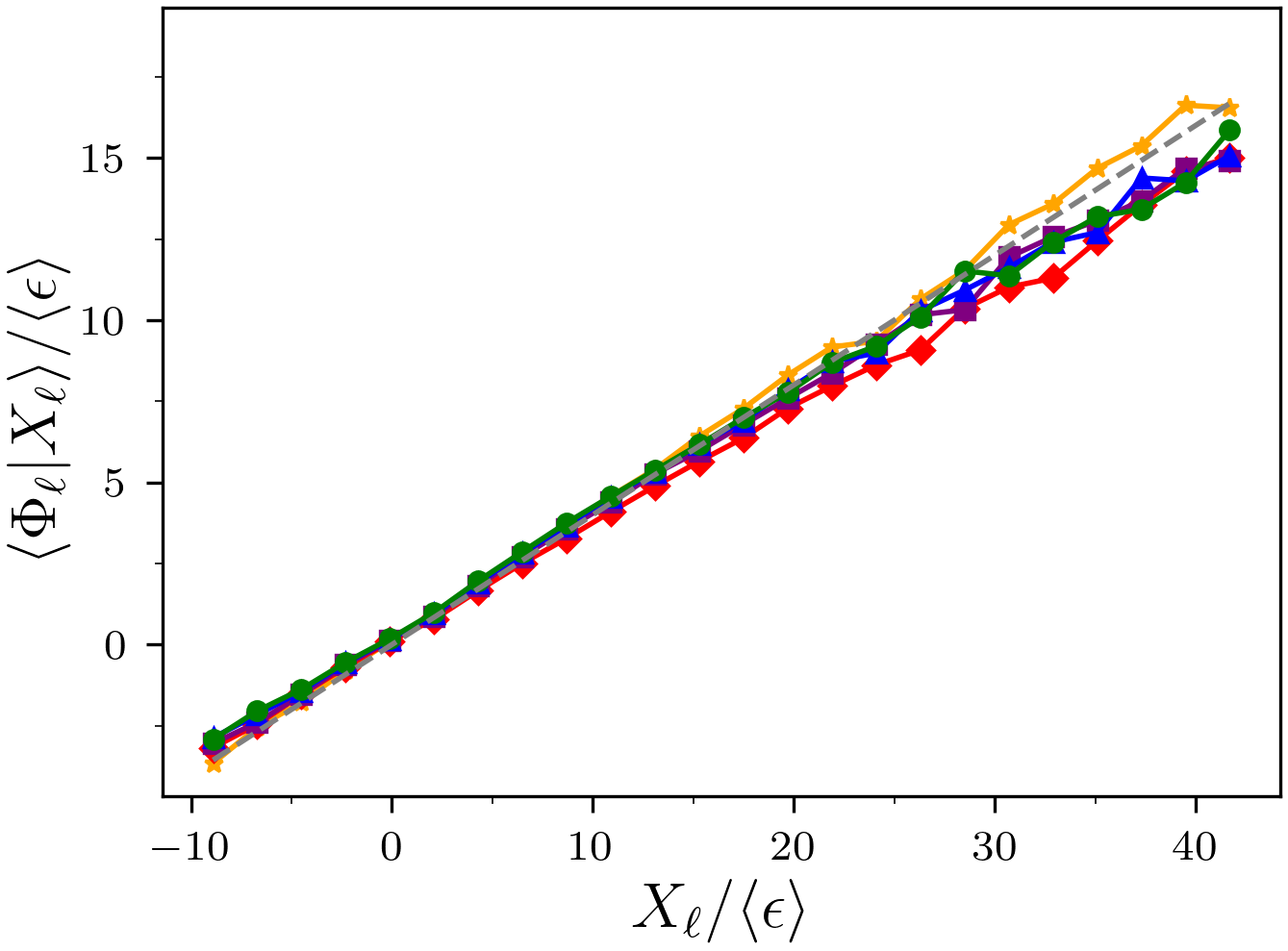}
    \caption{Conditional mean energy cascade rate
    $\langle\Phi_\ell\mid X_\ell\rangle/\langle\epsilon\rangle$
    as a function of $X_\ell/\langle\epsilon\rangle$, for $\ell/\eta=15$ (orange stars),
    $30$ (red diamonds), $45$ (purple squares), $60$
    (blue upward triangles), and $75$ (green circles). Dashed gray line show the theoretical prediction with slope $2/5$.}
    \label{fig:con_ave}
\end{figure}

To test validity of Eq.\eqref{tophat}, we define
$
    X_\ell
    \equiv
    (\frac{\ell}{2})^2\left(
    3\widetilde{R}_s+\frac{1}{4}\widetilde{V}_s
    \right)
$, so
Eq.~\eqref{tophat} can be written as
$
    \Phi_\ell
    =
    \frac{2}{5}X_\ell+O_\ell.
$
We first visualize the spatial distribution directly in a $550\times 550$ domain.
Figure~\ref{fig:Phi_SSAVS} compares an instantaneous planar
distribution of $\Phi_\ell$ with its VGT-resolved approximation $(2/5)X_\ell$ at $\ell=45\eta$. The two fields display similar spatial organization with elongated and filamentary morphologies of many intense forward-transfer (positive values) structures and localized inverse-transfer (negative values) regions occur at nearly the same positions. Differences in amplitude and in some of the smaller-scale features remain, as expected from the residual
$O_\ell$. We next quantify this correspondence statistically. For all five scales considered, $\ell/\eta=15$, $30$, $45$, $60$, and $75$, we find
$
    \left\langle (2/5)X_\ell\right\rangle
    \approx
    0.9\,\langle\Phi_\ell\rangle,
$
showing that the VGT-resolved contribution accounts for approximately
$90\%$ of the mean cascade rate. To determine whether it also captures
the dominant local statistical variation of $\Phi_\ell$, we condition
on $X_\ell$, such that
$
    \left\langle\Phi_\ell\mid X_\ell\right\rangle
    =
    \frac{2}{5}X_\ell
    +
    \left\langle O_\ell\mid X_\ell\right\rangle .
$
Figure~\ref{fig:con_ave} compares the measured conditional averages
with the theoretical prediction of slope $2/5$. Remarkably, the measured curves closely follow
the prediction at all five length scales from dissipative to inertial ranges, indicating
$
    \left\langle O_\ell\mid X_\ell\right\rangle\approx0
$
to leading statistical accuracy.

\begin{figure}[htbp]
    \hspace*{-0.8cm}
    \includegraphics[width=0.35\textwidth]{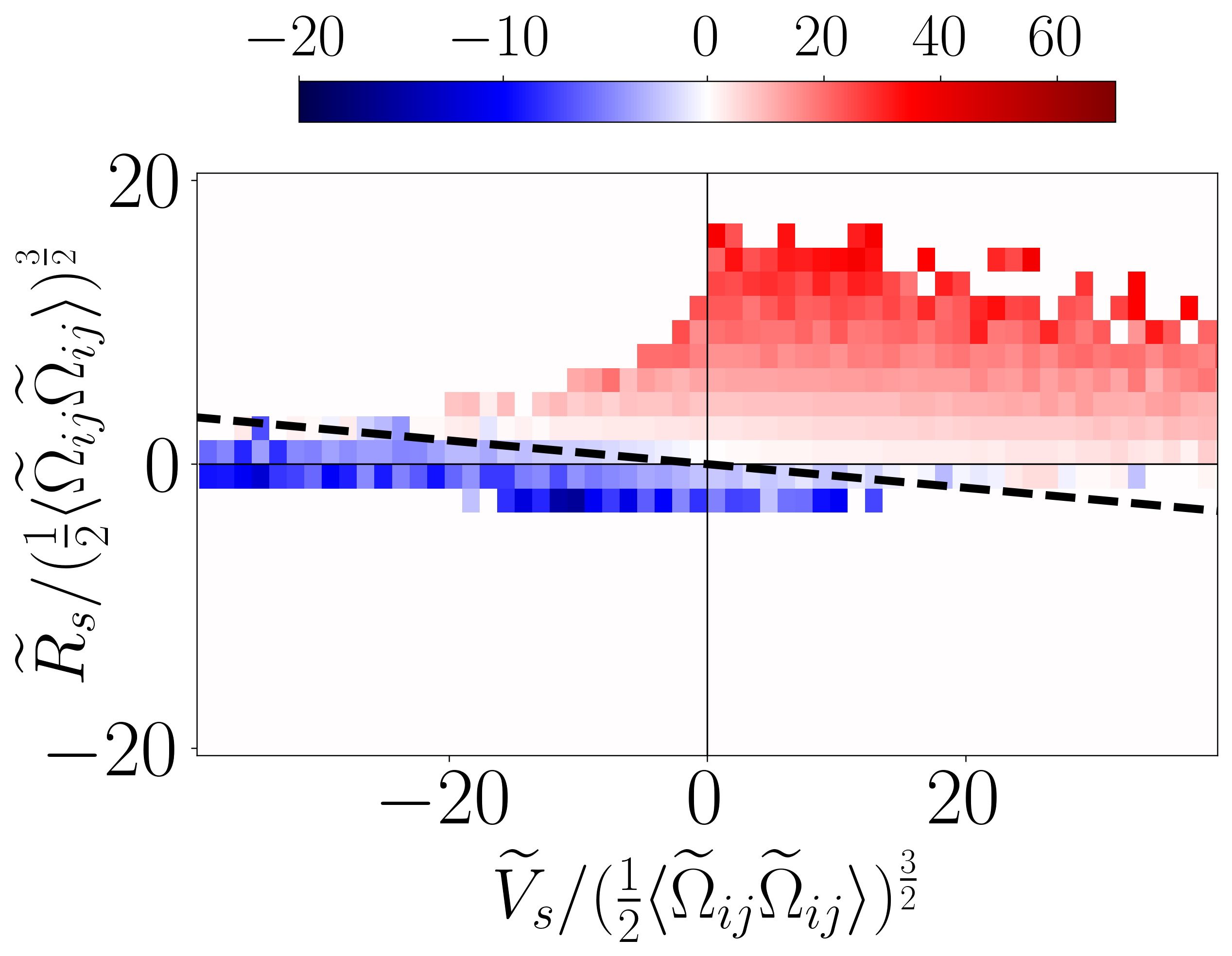 }
    \caption{Joint conditional average
    $\langle\Phi_\ell\mid\widetilde{R}_s,\widetilde{V}_s\rangle$ normalized by $\langle\epsilon\rangle$
    at $\ell=45\eta$ with $\widetilde{R}_s$ and $\widetilde{V}_s$ normalized by $\frac{1}{2}(\langle\widetilde\Omega_{ij}\widetilde\Omega_{ij}\rangle)^{\frac{3}{2}}$, where $\widetilde\Omega_{ij}=(\partial \tilde{u}_i/\partial x_j-\partial \tilde{u}_j/\partial x_i)/2$. Red and blue regions denote positive
    (forward) and negative (inverse) conditional mean energy transfer,
    respectively. The dashed line represents the analytical prediction of forward and inverse cascade, boundary $3\widetilde{R}_s+\widetilde{V}_s/4=0$.}
    \label{fig:combined_Phi_Pi}
\end{figure}

A more stringent test concerns the direction of the local cascade. We therefore consider the joint conditional average
$\langle\Phi_\ell\mid\widetilde{R}_s,\widetilde{V}_s\rangle$. Conditioning Eq.~\eqref{tophat} on the two VGT invariants (normalized by $(\frac{1}{2}\langle\widetilde\Omega_{ij}\widetilde\Omega_{ij}\rangle)^{\frac{3}{2}}$, where $\widetilde\Omega_{ij}=(\partial \tilde{u}_i/\partial x_j-\partial \tilde{u}_j/\partial x_i)/2$) gives
$
\langle\Phi_\ell\mid\widetilde{R}_s,\widetilde{V}_s\rangle
=
\frac{2}{5}r^2
(3\widetilde{R}_s+\widetilde{V}_s/4)
+
\langle O_\ell\mid\widetilde{R}_s,\widetilde{V}_s\rangle .
$
Hence, if the residual has little systematic dependence on
$(\widetilde{R}_s,\widetilde{V}_s)$ , the VGT
representation predicts a boundary of the conditional mean cascade along
$
3\widetilde{R}_s+\widetilde{V}_s/4=0.
$
Figure~\ref{fig:combined_Phi_Pi} shows the conditional statistics of $\Phi_\ell$ at $\ell=45\eta$. It is evident that the predicted line closely follows the observed boundary between positive and negative conditional mean
transfer. Regions on one side are predominantly associated with
$\langle\Phi_\ell\mid\widetilde{R}_s,\widetilde{V}_s\rangle>0$
(forward cascade), whereas those on the other side exhibit
$\langle\Phi_\ell\mid\widetilde{R}_s,\widetilde{V}_s\rangle<0$ (inverse cascade). Similar behaviour is observed at the other scales considered (see Supplemental Material). $3\widetilde{R}_s+\widetilde{V}_s/4=0$ thus
provides a simple statistical boundary separating
regions preferentially associated with forward and inverse energy transfer in the $(\widetilde{V}_s,\widetilde{R}_s)$ phase space. The close agreement of the predicted and measured sign boundaries also indicates that
$\langle O_\ell\mid\widetilde{R}_s,\widetilde{V}_s\rangle$
does not introduce a substantial systematic shift of the cascade direction.

Further physical insight is obtained by decomposing the local real-space flux into longitudinal and transverse contributions,
$
    \Phi_{\ell}
    =
    \Phi_{\ell}^{\parallel}
    +
    \Phi_{\ell}^{\perp},
$
where
$
    \Phi_{\ell}^{\parallel}
    =
    -\frac{3}{4\ell}\frac{1}{S_{\ell}}
    \oint_{S_{\ell}}
    (\delta\mathbf{u}\cdot\hat{\mathbf{n}})^3\,dS
$
and
$
    \Phi_{\ell}^{\perp}
    =
    -\frac{3}{4\ell}\frac{1}{S_{\ell}}
    \oint_{S_{\ell}}
    (\delta\mathbf{u}\cdot\hat{\mathbf{n}})
    \left\|
    \delta\mathbf{u}\times\hat{\mathbf{n}}
    \right\|^2\,dS .
$
This decomposition is exact. The longitudinal contribution can be
written as
$
    \Phi_{\ell}^{\parallel}
    =
    -3[\delta u_L^3]_{S_\ell}/(4\ell),
$
and is therefore directly connected to the longitudinal third-order
transfer underlying Kolmogorov's $4/5$ law,
$
    \left\langle[\delta u_L^3]_{S_\ell}\right\rangle
    \simeq
    -\frac{4}{5}\langle\epsilon\rangle\ell .
$

Applying the exact dipolar ($L=1$) component of the velocity
increment, Eq.~\eqref{du_est}, to the two channels separately gives
(see Supplemental Material)
\begin{align}
\Phi_{\ell}^{\parallel}(\mathbf{x},t)
&=
\frac{24}{35}
\left(\frac{\ell}{2}\right)^2
\widetilde{R}_s
+
O_\ell^{\parallel},
\label{para}\\
\Phi_{\ell}^{\perp}(\mathbf{x},t)
&=
\frac{18}{35}
\left(\frac{\ell}{2}\right)^2
\widetilde{R}_s
+
\frac{1}{10}
\left(\frac{\ell}{2}\right)^2
\widetilde{V}_s
+
O_\ell^{\perp}.
\label{trans}
\end{align}
Here $O_\ell^{\parallel}$ and $O_\ell^{\perp}$ contain all contributions involving the higher odd spherical-harmonic sectors $q_i$, including their nonlinear interactions with the dipolar
component.

Equations~\eqref{para} and \eqref{trans} reveal a striking separation of the underlying VGT dynamics. The VGT-resolved longitudinal transfer is governed solely by SSA, whereas VS enters only through the transverse channel. This distinction has a direct geometrical origin. For the dipolar velocity increment,
$
    \delta u_i^{(1)}
    =
    2r\widetilde A_{ij}\hat n_j
    =
    2r(\widetilde S_{ij}+\widetilde\Omega_{ij})\hat n_j,
$
the longitudinal component is
$
    \delta u_L^{(1)}
    =
    2r\,\hat n_i\widetilde A_{ij}\hat n_j .
$
Since $\widetilde\Omega_{ij}$ is antisymmetric,
$
    \hat n_i\widetilde\Omega_{ij}\hat n_j=0,
$
and therefore
$
    \delta u_L^{(1)}
    =
    2r\,\hat n_i\widetilde S_{ij}\hat n_j .
$
Thus rigid-body rotation generated by the antisymmetric part of the VGT produces no velocity increment normal to the spherical surface, i.e., it is entirely tangential. The longitudinal channel consequently samples only the strain field.

The transverse channel retains the tangential velocity and therefore
preserves the rotational part of the VGT. For the dipolar sector, the transverse component
$
    ||\delta\mathbf u_T^{(1)}||^2
    =
    4r^2
    \left[
    (\widetilde{A}_{ij}\hat{n}_j)^2
    -
    (\hat{n}_i\widetilde{S}_{ij}
    \hat{n}_j)^2
    \right].
$
Unlike the longitudinal increment,
$(\widetilde{A}_{ij}\hat{n}_j)^2$ contains both strain and
rotation. In particular,
$
    (\widetilde{\Omega}_{ij}\hat{n}_j)^2
    =
    [\widetilde\omega^2
    -
    (\widetilde{\omega}_i\hat{n}_i)^2]/4 .
$
When this transverse rotational motion is weighted by the radial strain
$\hat{n}_i\widetilde{S}_{ij}\hat{n}_j$ and
integrated over the sphere, spherical symmetry leaves the invariant
$
    \widetilde\omega_i
    \widetilde S_{ij}
    \widetilde\omega_j
    =
    \widetilde V_s .
$
Vortex stretching therefore enters the transverse transfer as the
coupling between tangential rotational motion and strain-induced radial
deformation. 


\begin{figure}[htbp] \centering \includegraphics[width=0.235\textwidth]{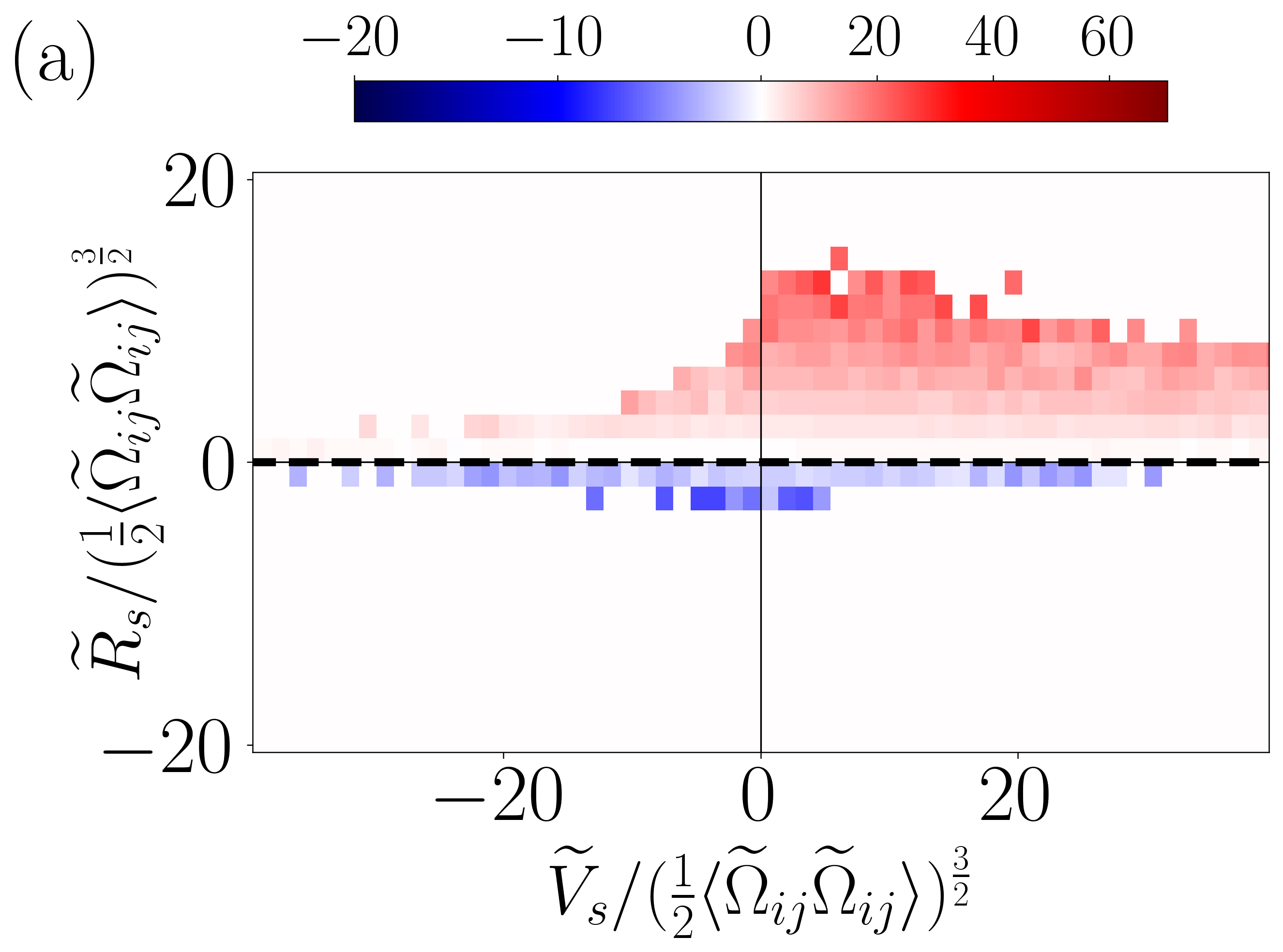} \hfill \includegraphics[width=0.235\textwidth]{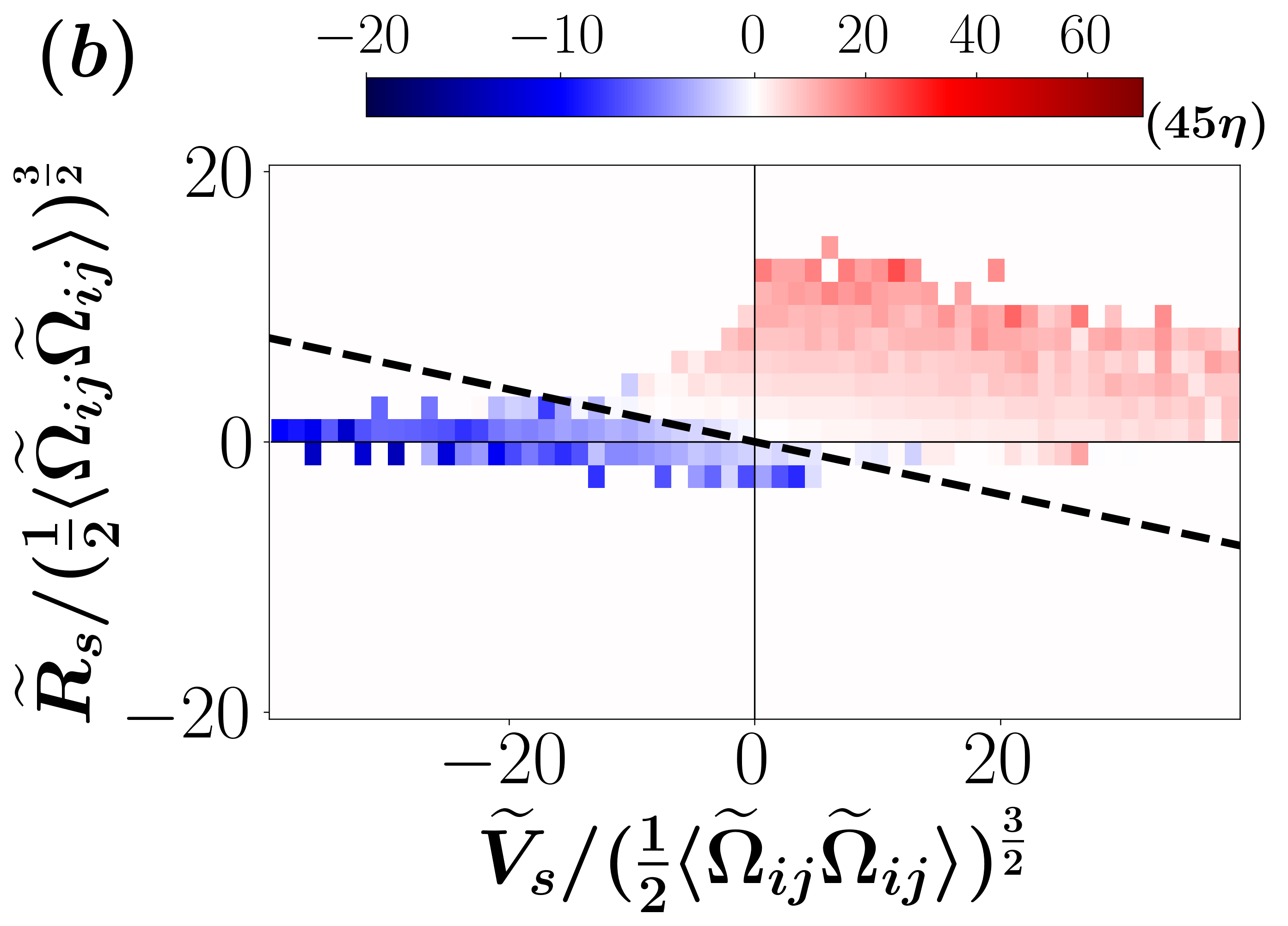} \caption{Joint conditional averages at $\ell=45\eta$ of (a) the longitudinal energy-transfer rate $\langle\Phi_{\ell}^{\parallel}\mid\widetilde{R}_s,\widetilde{V}_s\rangle$ and (b) the transverse energy-transfer rate $\langle\Phi_{\ell}^{\perp}\mid\widetilde{R}_s,\widetilde{V}_s\rangle$. Red and blue regions denote positive (forward) and negative (inverse) conditional mean transfer, respectively. The black dashed lines represent the analytical prediction of forward and inverse cascade boundary of the longitudinal and transverse energy fluxes shown in Eqs.~\eqref{para} and \eqref{trans}: $\widetilde{R}_s=0$ in (a) and $\frac{1}{10}\widetilde{V}_s+\frac{18}{35}\widetilde{R}_s=0$ in (b). Same normalization as shown in Fig. \ref{fig:combined_Phi_Pi}.} \label{fig:combined} \end{figure}

Equations~\eqref{para} and \eqref{trans} are tested through the joint
conditional averages
$\langle\Phi_{\ell}^{\parallel}\mid
\widetilde{R}_s,\widetilde{V}_s\rangle$
and
$\langle\Phi_{\ell}^{\perp}\mid
\widetilde{R}_s,\widetilde{V}_s\rangle$.
Figure~\ref{fig:combined} shows the results at $\ell=45\eta$.
For the longitudinal channel, Eq.~\eqref{para} predicts a
forward--inverse transition at $\widetilde{R}_s=0$, independent of
$\widetilde{V}_s$. This prediction closely follows the measured
zero-transfer boundary in Fig.~\ref{fig:combined}(a), consistent with
the geometrical result that the antisymmetric part of the VGT is
eliminated by the longitudinal projection. For the transverse component, Eq.~\eqref{trans} instead predicts
$
    \widetilde{V}_s/10
    +
    18\widetilde{R}_s/35
    =0,
$
which likewise closely tracks the measured sign transition in
Fig.~\ref{fig:combined}(b). Similar behavior is observed at the other scales considered (see Supplemental Material). The persistence of these predicted boundaries indicates that the higher-angular-mode residuals $O_\ell^{\parallel}$ and $O_\ell^{\perp}$ do not produce a substantial systematic displacement of the cascade direction in a statistical sense. We also observe an intenser region of negative conditional transfer in the transverse channel comparing panel (a) in Fig.~\ref{fig:combined}, suggesting that transverse velocity increments play a particularly important role in organizing local inverse transfer events. This information is absent from the classical $4/5$ law, which involves only the longitudinal third-order velocity increment.

A particularly direct consequence follows from the longitudinal
channel. Averaging Eq.~\eqref{para} and using Kolmogorov's $4/5$ law
gives
$
    \langle\Phi_{\ell}^{\parallel}\rangle
    =
    \frac{3}{5}\langle\epsilon\rangle
    =
    \frac{24}{35}r^2
    \langle\widetilde{R}_s\rangle
    +
    \langle O_\ell^{\parallel}\rangle ,
$
and therefore
$
    \langle\epsilon\rangle
    =
    \frac{8}{7}r^2
    \langle\widetilde{R}_s\rangle
    +
    \frac{5}{3}\langle O_\ell^{\parallel}\rangle .
$
Thus, when the mean higher-angular-mode contribution is subdominant, the $4/5$ law implies
$
    \langle\epsilon\rangle
    \simeq
    (8/7)r^2\langle\widetilde{R}_s\rangle
    =
    (2/7)\ell^2\langle\widetilde{R}_s\rangle .
$
This establishes a direct connection between Kolmogorov's
longitudinal third-order law and SSA of the coarse-grained VGT dynamics. Notably, it is consistent with the established role of SSA in sustaining the turbulent cascade and dissipation \cite{Alain}.

We finally turn to the mechanical meaning of the geometrical
coefficient appearing in Eq.~\eqref{tophat}. Defining
\begin{equation}
    I
    \equiv
    \frac{2}{5}
    \left(\frac{\ell}{2}\right)^2,
    \qquad
    \Phi_{\ell}(\mathbf{x},t)
    =
    I
    \left(
    3\widetilde R_s
    +
    \frac14\widetilde V_s
    \right)
    +
    O_\ell ,
\label{Phi_inertia}
\end{equation}
we recognize $I$ as the moment of inertia per unit mass of a uniform sphere of radius $r=\ell/2$ about an axis through its center. This identification suggests a direct mechanical interpretation of the VGT combination governing the local scale-space energy flux.

Consider first the restricted-Euler dynamics
\cite{meneveau2011lagrangian} (we use the same VGT notation as above, but without tildes). For a spherical element with specific
moment of inertia $I$, we define the rotational and deformational
kinetic energies as
$
E_{\rm rot}=I\omega_i\omega_i/8
$
and
$
E_{\rm def}=IS_{ij}S_{ij}/4,
$
respectively. The former is the usual rigid-body rotational energy,
$
E_{\rm rot}=I|\boldsymbol{\omega}/2|^2/2,
$
while the latter is the kinetic energy associated with linear
deformation of a uniform spherical element. The vorticity equation
$
d\omega_i/dt=S_{ij}\omega_j
$
gives
$
    dE_{\rm rot}/dt
    =
    IV_s/4,
$
showing that vortex stretching directly controls the production or
depletion of rotational kinetic energy. For the deformational part,
defining
$
Q_s=-S_{ij}S_{ij}/2,
$
with
$
dQ_s/dt=-3R_s+V_s/4,
$
yields
$
    dE_{\rm def}/dt
    =
    3IR_s/2
    -
    IV_s/8.
$
Thus SSA acts directly on the deformational energy, whereas VS couples
the rotational and deformational motions, i.e., a positive $V_s$ increases
the rotational energy while contributing with the opposite sign to the
deformational energy. Combining the two energy-rate equations gives
\begin{equation}
    \frac{d}{dt}
    \left[
    2\left(E_{\rm rot}+E_{\rm def}\right)
    \right]
    =
    I
    \left(
    3R_s+\frac14V_s
    \right).
\label{eq:7}
\end{equation}
Hence, within restricted-Euler dynamics,
$I(3R_s+V_s/4)$ is exactly twice the rate of change of the combined
rotational and deformational kinetic energy of a spherical fluid
element.

The corresponding relation for the coarse-grained VGT follows from its exact transport equation,$
    \widetilde D\widetilde A_{ij} / Dt
    =
    -\widetilde A_{ik}\widetilde A_{kj}
    +
    \frac13\delta_{ij}
    \widetilde A_{mn}\widetilde A_{nm}
    +
    \widetilde M_{ij},$
where $\widetilde M_{ij}$ collects the anisotropic pressure-Hessian, SGS-stress, viscous, and forcing contributions
\cite{Johnson2024annual}. Defining the corresponding coarse-grained
mechanical energies,
$
\widetilde E_{\rm rot}
=
I\widetilde\omega_i\widetilde\omega_i/8
$
and
$
\widetilde E_{\rm def}
=
I\widetilde S_{ij}\widetilde S_{ij}/4,
$
and holding the scale $\ell$, and hence $I$, fixed, contraction of the
VGT equation gives
\begin{equation}
    \frac{\widetilde D}{Dt}
    \left[
    2\left(
    \widetilde E_{\rm rot}
    +
    \widetilde E_{\rm def}
    \right)
    \right]
    =
    I
    \left(
    3\widetilde R_s
    +
    \frac14\widetilde V_s
    \right)
    +
    I\widetilde A_{ij}\widetilde M_{ij}.
\label{eq:filtered_energy}
\end{equation}
The first term on the right of Eq.~\eqref{eq:filtered_energy} is
generated by the local nonlinear self-interaction of the
coarse-grained VGT and is precisely the VGT-resolved contribution
appearing in Eq.~\eqref{Phi_inertia}. Comparison of
Eqs.~\eqref{Phi_inertia}, \eqref{eq:7}, and
\eqref{eq:filtered_energy} therefore reveals a direct mechanical
correspondence, i.e., the VGT-resolved contribution to the scale-space energy flux is exactly the same nonlinear term that governs the rate of change of twice the rotational and deformational mechanical energy at scale $\ell$. 




In summary, we have established a geometrical route from two-point velocity increments to local coarse-grained velocity-gradient dynamics. The dominant VGT-resolved contribution to the real-space turbulent energy cascade is governed locally by strain self-amplification and vortex stretching, and is mechanically linked to the evolution of the rotational and deformational energy of a spherical fluid neighborhood.
This connection further yields simple statistical boundaries separating forward and inverse transfer for the total flux and for its longitudinal and transverse components. This provides a new mechanical perspective and mechanical picture of turbulence energy cascade, in which the direction of interscale energy transfer emerges from the competition between the nonlinear VGT mechanisms associated with rotational and deformational fluid element dynamics.

Importantly, this mechanical picture and the correspondence of Eq.\eqref{tophat} are not tied to a particular scale
$\ell$ at dissipative or inertial range, i.e., changing $\ell$ modifies the filtered VGT and the relative importance of the residual, but not the identification of the dipolar mode or the SSA and VS mechanisms associated with it. This scale-independent geometrical structure also clarifies the scope of the present
description. The tensor $\widetilde A_{ij}$ contains the complete local rank-two information of the $L=1$ sector, whereas the residual $q_i$ contains genuinely higher angular modes that cannot be represented by
the VGT alone. Understanding the dynamics of these higher sectors, their coupling to the dipolar mode , and whether they encode nonlocal interactions (e.g, relation to $I\widetilde A_{ij}\widetilde M_{ij}$ \cite{johnson2020local}) in scale or space therefore provides a natural future work. 

More broadly, the present work establishes a framework for connecting spherical two-point statistics to local coarse-grained dynamics. The same strategy may be extended to other quadratic invariants and cascade processes, including helicity in three-dimensional turbulence, energy and enstrophy in two-dimensional turbulence, scalar variance in advected scalar fields, and magnetic and Els\"asser-energy transfer in magnetohydrodynamic (MHD) turbulence. It may also provide a useful starting point for anisotropic and inhomogeneous flows, where higher spherical-harmonic sectors can carry additional directional information. For turbulence in spherical or curved geometries, such as
atmospheric and oceanic flows on planetary surfaces or convection and MHD in spherical shells, analogous constructions based on vector spherical harmonics and appropriately defined coarse-grained velocity gradients may provide a route from two-point statistics to local dynamical mechanisms.

{\bf Acknowledgements:} The authors thank Professors: C. Meneveau, A. Pumir and X. Lu for valuable comments and suggestions; and thank JHTDB/IDIES for providing data. This work is supported by the  National Natural Science Foundation of China (grant
No. 12388101)  

\bibliographystyle{apsrev4-2}
\bibliography{Resubmission_folder_Febr2024/2D_PRL_2026} 

\begin{thebibliography}{24}%
\makeatletter
\providecommand \@ifxundefined [1]{%
 \@ifx{#1\undefined}
}%
\providecommand \@ifnum [1]{%
 \ifnum #1\expandafter \@firstoftwo
 \else \expandafter \@secondoftwo
 \fi
}%
\providecommand \@ifx [1]{%
 \ifx #1\expandafter \@firstoftwo
 \else \expandafter \@secondoftwo
 \fi
}%
\providecommand \natexlab [1]{#1}%
\providecommand \enquote  [1]{``#1''}%
\providecommand \bibnamefont  [1]{#1}%
\providecommand \bibfnamefont [1]{#1}%
\providecommand \citenamefont [1]{#1}%
\providecommand \href@noop [0]{\@secondoftwo}%
\providecommand \href [0]{\begingroup \@sanitize@url \@href}%
\providecommand \@href[1]{\@@startlink{#1}\@@href}%
\providecommand \@@href[1]{\endgroup#1\@@endlink}%
\providecommand \@sanitize@url [0]{\catcode `\\12\catcode `\$12\catcode `\&12\catcode `\#12\catcode `\^12\catcode `\_12\catcode `\%12\relax}%
\providecommand \@@startlink[1]{}%
\providecommand \@@endlink[0]{}%
\providecommand \url  [0]{\begingroup\@sanitize@url \@url }%
\providecommand \@url [1]{\endgroup\@href {#1}{\urlprefix }}%
\providecommand \urlprefix  [0]{URL }%
\providecommand \Eprint [0]{\href }%
\providecommand \doibase [0]{https://doi.org/}%
\providecommand \selectlanguage [0]{\@gobble}%
\providecommand \bibinfo  [0]{\@secondoftwo}%
\providecommand \bibfield  [0]{\@secondoftwo}%
\providecommand \translation [1]{[#1]}%
\providecommand \BibitemOpen [0]{}%
\providecommand \bibitemStop [0]{}%
\providecommand \bibitemNoStop [0]{.\EOS\space}%
\providecommand \EOS [0]{\spacefactor3000\relax}%
\providecommand \BibitemShut  [1]{\csname bibitem#1\endcsname}%
\let\auto@bib@innerbib\@empty
\bibitem [{\citenamefont {Richardson}(1922)}]{Richardson1922}%
  \BibitemOpen
  \bibfield  {author} {\bibinfo {author} {\bibfnamefont {L.}~\bibnamefont {Richardson}},\ }\href@noop {} {\  (\bibinfo {year} {1922})}\BibitemShut {NoStop}%
\bibitem [{\citenamefont {Kolmogorov}(1941)}]{K41}%
  \BibitemOpen
  \bibfield  {author} {\bibinfo {author} {\bibfnamefont {A.~N.}\ \bibnamefont {Kolmogorov}},\ }\href@noop {} {\bibfield  {journal} {\bibinfo  {journal} {CR Acad. Sci. URSS}\ }\textbf {\bibinfo {volume} {30}},\ \bibinfo {pages} {301} (\bibinfo {year} {1941})}\BibitemShut {NoStop}%
\bibitem [{\citenamefont {Piomelli}\ \emph {et~al.}(1991)\citenamefont {Piomelli}, \citenamefont {Cabot}, \citenamefont {Moin},\ and\ \citenamefont {Lee}}]{Piomelli1991}%
  \BibitemOpen
  \bibfield  {author} {\bibinfo {author} {\bibfnamefont {U.}~\bibnamefont {Piomelli}}, \bibinfo {author} {\bibfnamefont {W.~H.}\ \bibnamefont {Cabot}}, \bibinfo {author} {\bibfnamefont {P.}~\bibnamefont {Moin}},\ and\ \bibinfo {author} {\bibfnamefont {S.}~\bibnamefont {Lee}},\ }\href@noop {} {\bibfield  {journal} {\bibinfo  {journal} {Physics of Fluids A}\ }\textbf {\bibinfo {volume} {3}},\ \bibinfo {pages} {1766} (\bibinfo {year} {1991})}\BibitemShut {NoStop}%
\bibitem [{\citenamefont {Meneveau}\ and\ \citenamefont {Katz}(2000)}]{Meneveau2000}%
  \BibitemOpen
  \bibfield  {author} {\bibinfo {author} {\bibfnamefont {C.}~\bibnamefont {Meneveau}}\ and\ \bibinfo {author} {\bibfnamefont {J.}~\bibnamefont {Katz}},\ }\href {https://doi.org/10.1146/annurev.fluid.32.1.1} {\bibfield  {journal} {\bibinfo  {journal} {Annual Review of Fluid Mechanics}\ }\textbf {\bibinfo {volume} {32}},\ \bibinfo {pages} {1} (\bibinfo {year} {2000})}\BibitemShut {NoStop}%
\bibitem [{\citenamefont {Borue}\ and\ \citenamefont {Orszag}(1998)}]{borue1998local}%
  \BibitemOpen
  \bibfield  {author} {\bibinfo {author} {\bibfnamefont {V.}~\bibnamefont {Borue}}\ and\ \bibinfo {author} {\bibfnamefont {S.~A.}\ \bibnamefont {Orszag}},\ }\href@noop {} {\bibfield  {journal} {\bibinfo  {journal} {Journal of Fluid Mechanics}\ }\textbf {\bibinfo {volume} {366}},\ \bibinfo {pages} {1} (\bibinfo {year} {1998})}\BibitemShut {NoStop}%
\bibitem [{\citenamefont {Yao}\ \emph {et~al.}(2024{\natexlab{a}})\citenamefont {Yao}, \citenamefont {Schnaubelt}, \citenamefont {Szalay}, \citenamefont {Zaki},\ and\ \citenamefont {Meneveau}}]{yao2024comparing}%
  \BibitemOpen
  \bibfield  {author} {\bibinfo {author} {\bibfnamefont {H.}~\bibnamefont {Yao}}, \bibinfo {author} {\bibfnamefont {M.}~\bibnamefont {Schnaubelt}}, \bibinfo {author} {\bibfnamefont {A.~S.}\ \bibnamefont {Szalay}}, \bibinfo {author} {\bibfnamefont {T.~A.}\ \bibnamefont {Zaki}},\ and\ \bibinfo {author} {\bibfnamefont {C.}~\bibnamefont {Meneveau}},\ }\href@noop {} {\bibfield  {journal} {\bibinfo  {journal} {Journal of Fluid Mechanics}\ }\textbf {\bibinfo {volume} {980}},\ \bibinfo {pages} {A42} (\bibinfo {year} {2024}{\natexlab{a}})}\BibitemShut {NoStop}%
\bibitem [{\citenamefont {Taylor}(1938)}]{taylor1938production}%
  \BibitemOpen
  \bibfield  {author} {\bibinfo {author} {\bibfnamefont {G.~I.}\ \bibnamefont {Taylor}},\ }\href@noop {} {\bibfield  {journal} {\bibinfo  {journal} {Proceedings of the Royal Society of London A}\ }\textbf {\bibinfo {volume} {164}},\ \bibinfo {pages} {15} (\bibinfo {year} {1938})}\BibitemShut {NoStop}%
\bibitem [{\citenamefont {Onsager}(1949)}]{onsager1949statistical}%
  \BibitemOpen
  \bibfield  {author} {\bibinfo {author} {\bibfnamefont {L.}~\bibnamefont {Onsager}},\ }\href@noop {} {\bibfield  {journal} {\bibinfo  {journal} {Il Nuovo Cimento (1943--1954)}\ }\textbf {\bibinfo {volume} {6}},\ \bibinfo {pages} {279} (\bibinfo {year} {1949})}\BibitemShut {NoStop}%
\bibitem [{\citenamefont {Tennekes}\ and\ \citenamefont {Lumley}(1972)}]{tennekes1972first}%
  \BibitemOpen
  \bibfield  {author} {\bibinfo {author} {\bibfnamefont {H.}~\bibnamefont {Tennekes}}\ and\ \bibinfo {author} {\bibfnamefont {J.~L.}\ \bibnamefont {Lumley}},\ }\href@noop {} {\emph {\bibinfo {title} {A First Course in Turbulence}}}\ (\bibinfo  {publisher} {MIT Press},\ \bibinfo {address} {Cambridge, MA},\ \bibinfo {year} {1972})\BibitemShut {NoStop}%
\bibitem [{\citenamefont {Germano}(1992)}]{germano1992turbulence}%
  \BibitemOpen
  \bibfield  {author} {\bibinfo {author} {\bibfnamefont {M.}~\bibnamefont {Germano}},\ }\href@noop {} {\bibfield  {journal} {\bibinfo  {journal} {Journal of Fluid Mechanics}\ }\textbf {\bibinfo {volume} {238}},\ \bibinfo {pages} {325} (\bibinfo {year} {1992})}\BibitemShut {NoStop}%
\bibitem [{\citenamefont {Johnson}(2020)}]{johnson2020local}%
  \BibitemOpen
  \bibfield  {author} {\bibinfo {author} {\bibfnamefont {P.~L.}\ \bibnamefont {Johnson}},\ }\href@noop {} {\bibfield  {journal} {\bibinfo  {journal} {Physical Review Letters}\ }\textbf {\bibinfo {volume} {124}},\ \bibinfo {pages} {104501} (\bibinfo {year} {2020})}\BibitemShut {NoStop}%
\bibitem [{\citenamefont {Portela}\ \emph {et~al.}(2017)\citenamefont {Portela}, \citenamefont {Papadakis},\ and\ \citenamefont {Vassilicos}}]{portela2017turbulence}%
  \BibitemOpen
  \bibfield  {author} {\bibinfo {author} {\bibfnamefont {F.~A.}\ \bibnamefont {Portela}}, \bibinfo {author} {\bibfnamefont {G.}~\bibnamefont {Papadakis}},\ and\ \bibinfo {author} {\bibfnamefont {J.~C.}\ \bibnamefont {Vassilicos}},\ }\href@noop {} {\bibfield  {journal} {\bibinfo  {journal} {Journal of Fluid Mechanics}\ }\textbf {\bibinfo {volume} {825}},\ \bibinfo {pages} {315} (\bibinfo {year} {2017})}\BibitemShut {NoStop}%
\bibitem [{\citenamefont {Hill}(2002)}]{hill2002exact}%
  \BibitemOpen
  \bibfield  {author} {\bibinfo {author} {\bibfnamefont {R.~J.}\ \bibnamefont {Hill}},\ }\href@noop {} {\bibfield  {journal} {\bibinfo  {journal} {Journal of Fluid Mechanics}\ }\textbf {\bibinfo {volume} {468}},\ \bibinfo {pages} {317} (\bibinfo {year} {2002})}\BibitemShut {NoStop}%
\bibitem [{\citenamefont {Yao}\ \emph {et~al.}(2023)\citenamefont {Yao}, \citenamefont {Zaki},\ and\ \citenamefont {Meneveau}}]{yao2023entropy}%
  \BibitemOpen
  \bibfield  {author} {\bibinfo {author} {\bibfnamefont {H.}~\bibnamefont {Yao}}, \bibinfo {author} {\bibfnamefont {T.~A.}\ \bibnamefont {Zaki}},\ and\ \bibinfo {author} {\bibfnamefont {C.}~\bibnamefont {Meneveau}},\ }\href@noop {} {\bibfield  {journal} {\bibinfo  {journal} {Journal of Fluid Mechanics}\ }\textbf {\bibinfo {volume} {973}},\ \bibinfo {pages} {R6} (\bibinfo {year} {2023})}\BibitemShut {NoStop}%
\bibitem [{\citenamefont {Yao}\ \emph {et~al.}(2024{\natexlab{b}})\citenamefont {Yao}, \citenamefont {Yeung}, \citenamefont {Zaki},\ and\ \citenamefont {Meneveau}}]{yao2024forward}%
  \BibitemOpen
  \bibfield  {author} {\bibinfo {author} {\bibfnamefont {H.}~\bibnamefont {Yao}}, \bibinfo {author} {\bibfnamefont {P.}~\bibnamefont {Yeung}}, \bibinfo {author} {\bibfnamefont {T.}~\bibnamefont {Zaki}},\ and\ \bibinfo {author} {\bibfnamefont {C.}~\bibnamefont {Meneveau}},\ }\href@noop {} {\bibfield  {journal} {\bibinfo  {journal} {Physical review letters}\ }\textbf {\bibinfo {volume} {132}},\ \bibinfo {pages} {164001} (\bibinfo {year} {2024}{\natexlab{b}})}\BibitemShut {NoStop}%
\bibitem [{\citenamefont {Carbone}\ and\ \citenamefont {Bragg}(2020)}]{carbone2020vortex}%
  \BibitemOpen
  \bibfield  {author} {\bibinfo {author} {\bibfnamefont {M.}~\bibnamefont {Carbone}}\ and\ \bibinfo {author} {\bibfnamefont {A.~D.}\ \bibnamefont {Bragg}},\ }\href {https://doi.org/10.1017/jfm.2019.923} {\bibfield  {journal} {\bibinfo  {journal} {Journal of Fluid Mechanics}\ }\textbf {\bibinfo {volume} {883}},\ \bibinfo {pages} {R2} (\bibinfo {year} {2020})}\BibitemShut {NoStop}%
\bibitem [{\citenamefont {Arad}\ \emph {et~al.}(1998)\citenamefont {Arad}, \citenamefont {Dhruva}, \citenamefont {Kurien}, \citenamefont {L'vov}, \citenamefont {Procaccia},\ and\ \citenamefont {Sreenivasan}}]{Arad1998}%
  \BibitemOpen
  \bibfield  {author} {\bibinfo {author} {\bibfnamefont {I.}~\bibnamefont {Arad}}, \bibinfo {author} {\bibfnamefont {B.}~\bibnamefont {Dhruva}}, \bibinfo {author} {\bibfnamefont {S.}~\bibnamefont {Kurien}}, \bibinfo {author} {\bibfnamefont {V.~S.}\ \bibnamefont {L'vov}}, \bibinfo {author} {\bibfnamefont {I.}~\bibnamefont {Procaccia}},\ and\ \bibinfo {author} {\bibfnamefont {K.~R.}\ \bibnamefont {Sreenivasan}},\ }\href {https://doi.org/10.1103/PhysRevLett.81.5330} {\bibfield  {journal} {\bibinfo  {journal} {Phys. Rev. Lett.}\ }\textbf {\bibinfo {volume} {81}},\ \bibinfo {pages} {5330} (\bibinfo {year} {1998})}\BibitemShut {NoStop}%
\bibitem [{\citenamefont {Arad}\ \emph {et~al.}(1999{\natexlab{a}})\citenamefont {Arad}, \citenamefont {Biferale}, \citenamefont {Mazzitelli},\ and\ \citenamefont {Procaccia}}]{Arad1999a}%
  \BibitemOpen
  \bibfield  {author} {\bibinfo {author} {\bibfnamefont {I.}~\bibnamefont {Arad}}, \bibinfo {author} {\bibfnamefont {L.}~\bibnamefont {Biferale}}, \bibinfo {author} {\bibfnamefont {I.}~\bibnamefont {Mazzitelli}},\ and\ \bibinfo {author} {\bibfnamefont {I.}~\bibnamefont {Procaccia}},\ }\href {https://doi.org/10.1103/PhysRevLett.82.5040} {\bibfield  {journal} {\bibinfo  {journal} {Phys. Rev. Lett.}\ }\textbf {\bibinfo {volume} {82}},\ \bibinfo {pages} {5040} (\bibinfo {year} {1999}{\natexlab{a}})}\BibitemShut {NoStop}%
\bibitem [{\citenamefont {Arad}\ \emph {et~al.}(1999{\natexlab{b}})\citenamefont {Arad}, \citenamefont {L'vov},\ and\ \citenamefont {Procaccia}}]{Arad1999b}%
  \BibitemOpen
  \bibfield  {author} {\bibinfo {author} {\bibfnamefont {I.}~\bibnamefont {Arad}}, \bibinfo {author} {\bibfnamefont {V.~S.}\ \bibnamefont {L'vov}},\ and\ \bibinfo {author} {\bibfnamefont {I.}~\bibnamefont {Procaccia}},\ }\href {https://doi.org/10.1103/PhysRevE.59.6753} {\bibfield  {journal} {\bibinfo  {journal} {Phys. Rev. E}\ }\textbf {\bibinfo {volume} {59}},\ \bibinfo {pages} {6753} (\bibinfo {year} {1999}{\natexlab{b}})}\BibitemShut {NoStop}%
\bibitem [{\citenamefont {Kurien}\ \emph {et~al.}(2001)\citenamefont {Kurien}, \citenamefont {Aivalis},\ and\ \citenamefont {Sreenivasan}}]{Kurien2001}%
  \BibitemOpen
  \bibfield  {author} {\bibinfo {author} {\bibfnamefont {S.}~\bibnamefont {Kurien}}, \bibinfo {author} {\bibfnamefont {K.~G.}\ \bibnamefont {Aivalis}},\ and\ \bibinfo {author} {\bibfnamefont {K.~R.}\ \bibnamefont {Sreenivasan}},\ }\href {https://doi.org/10.1017/S0022112001005913} {\bibfield  {journal} {\bibinfo  {journal} {J. Fluid Mech.}\ }\textbf {\bibinfo {volume} {448}},\ \bibinfo {pages} {279} (\bibinfo {year} {2001})}\BibitemShut {NoStop}%
\bibitem [{\citenamefont {Yao}\ \emph {et~al.}(2025)\citenamefont {Yao}, \citenamefont {Schnaubelt}, \citenamefont {Lubonja}, \citenamefont {Medvedev}, \citenamefont {Hao}, \citenamefont {Wang}, \citenamefont {Lemson}, \citenamefont {Burns}, \citenamefont {Szalay}, \citenamefont {Yeung} \emph {et~al.}}]{yao2025analysis}%
  \BibitemOpen
  \bibfield  {author} {\bibinfo {author} {\bibfnamefont {H.}~\bibnamefont {Yao}}, \bibinfo {author} {\bibfnamefont {M.}~\bibnamefont {Schnaubelt}}, \bibinfo {author} {\bibfnamefont {A.}~\bibnamefont {Lubonja}}, \bibinfo {author} {\bibfnamefont {D.}~\bibnamefont {Medvedev}}, \bibinfo {author} {\bibfnamefont {Y.}~\bibnamefont {Hao}}, \bibinfo {author} {\bibfnamefont {M.}~\bibnamefont {Wang}}, \bibinfo {author} {\bibfnamefont {G.}~\bibnamefont {Lemson}}, \bibinfo {author} {\bibfnamefont {R.}~\bibnamefont {Burns}}, \bibinfo {author} {\bibfnamefont {A.}~\bibnamefont {Szalay}}, \bibinfo {author} {\bibfnamefont {P.}~\bibnamefont {Yeung}}, \emph {et~al.},\ }\href@noop {} {\bibfield  {journal} {\bibinfo  {journal} {Journal of Turbulence}\ }\textbf {\bibinfo {volume} {26}},\ \bibinfo {pages} {273} (\bibinfo {year} {2025})}\BibitemShut {NoStop}%
\bibitem [{\citenamefont {Buaria}\ \emph {et~al.}(2022)\citenamefont {Buaria}, \citenamefont {Pumir},\ and\ \citenamefont {Bodenschatz}}]{Alain}%
  \BibitemOpen
  \bibfield  {author} {\bibinfo {author} {\bibfnamefont {D.}~\bibnamefont {Buaria}}, \bibinfo {author} {\bibfnamefont {A.}~\bibnamefont {Pumir}},\ and\ \bibinfo {author} {\bibfnamefont {E.}~\bibnamefont {Bodenschatz}},\ }\href {https://doi.org/10.1098/rsta.2021.0088} {\bibfield  {journal} {\bibinfo  {journal} {Philosophical Transactions of the Royal Society A: Mathematical, Physical and Engineering Sciences}\ }\textbf {\bibinfo {volume} {380}},\ \bibinfo {pages} {20210088} (\bibinfo {year} {2022})},\ \Eprint {https://arxiv.org/abs/2107.01719} {arXiv:2107.01719 [physics.flu-dyn]} \BibitemShut {NoStop}%
\bibitem [{\citenamefont {Meneveau}(2011)}]{meneveau2011lagrangian}%
  \BibitemOpen
  \bibfield  {author} {\bibinfo {author} {\bibfnamefont {C.}~\bibnamefont {Meneveau}},\ }\href {https://doi.org/10.1146/annurev-fluid-122109-160708} {\bibfield  {journal} {\bibinfo  {journal} {Annual Review of Fluid Mechanics}\ }\textbf {\bibinfo {volume} {43}},\ \bibinfo {pages} {219} (\bibinfo {year} {2011})}\BibitemShut {NoStop}%
\bibitem [{\citenamefont {Johnson}\ and\ \citenamefont {Wilczek}(2024)}]{Johnson2024annual}%
  \BibitemOpen
  \bibfield  {author} {\bibinfo {author} {\bibfnamefont {P.~L.}\ \bibnamefont {Johnson}}\ and\ \bibinfo {author} {\bibfnamefont {M.}~\bibnamefont {Wilczek}},\ }\href {https://doi.org/10.1146/annurev-fluid-032823-035457} {\bibfield  {journal} {\bibinfo  {journal} {Annual Review of Fluid Mechanics}\ }\textbf {\bibinfo {volume} {56}},\ \bibinfo {pages} {457} (\bibinfo {year} {2024})}\BibitemShut {NoStop}%
\end{thebibliography}%


\end{document}


\preprint{APS/123-QED}

\title{Supplemental Material:\\``Geometric Mechanics of Real-Space Energy Transfer in Turbulence''}

\author{Weijie, Zhang}
\affiliation{
 Department of Modern Mechanics, University of Science and Technology of China, Hefei, China
} 

\author{Hanxun, Yao}
\affiliation{
 Department of Modern Mechanics, University of Science and Technology of China, Hefei, China
}

\date{\today} 
\begin{abstract}
This Supplemental Material provides detailed derivations and additional conditional statistics supporting the main Letter. We first derive the spherical-harmonic decomposition of the finite velocity increment and its exact connection to the filtered velocity-gradient tensor (VGT). We then obtain the $L=1$ VGT-resolved contribution to the local real-space energy flux through an eigenvalue decomposition, followed by the corresponding derivation for its longitudinal and transverse
orthogonal components. Additional joint conditional statistics at four scales, $\ell/\eta=15$, $30$, $60$, and $75$, spanning the
viscous dissipative to inertial ranges, demonstrate the scale robustness
of the predicted forward and inverse transfer boundaries.
\end{abstract}
\maketitle

\section{Spherical harmonic representation of the velocity increment}

In this section, we establish a direct relation between the finite velocity increment
$
\delta u_i(\mathbf{x},\mathbf r)
=
u_i(\mathbf{x}+\mathbf r)-u_i(\mathbf{x}-\mathbf r)
$
and the filtered velocity-gradient tensor
$\widetilde A_{ij}=\partial_j\widetilde u_i$ for a spherical top-hat
filter. With radius $r=\ell/2$, the filtered velocity is
\begin{equation}
    \widetilde u_i(\mathbf{x})
    =
    \frac{1}{V_\ell}
    \int_{V_\ell}
    u_i(\mathbf{x}+\mathbf r)\,d^3\mathbf r,
\end{equation}
where $V_\ell=4\pi r^3/3$. Since the
integration domain is fixed in the relative coordinate $\mathbf r$ (independent to $\mathbf{x}$ ),
\begin{equation}
    \widetilde A_{ij}
    =
    \partial_j\widetilde u_i
    =
    \frac{1}{V_\ell}
    \int_{V_\ell}
    \partial_j u_i(\mathbf{x}+\mathbf r)\,d^3\mathbf r .
\end{equation}
Introducing $\mathbf y=\mathbf x+\mathbf r$, the velocity
$u_i(\mathbf x+\mathbf r)=u_i(\mathbf y)$ depends on
$\mathbf x$ and $\mathbf r$ only through their sum. The chain rule
therefore gives
$
\partial_{x_j}u_i(\mathbf x+\mathbf r)
=
(\partial u_i/\partial y_k)\delta_{kj}
=
\partial_{r_j}u_i(\mathbf x+\mathbf r).
$
Consequently, $\partial/\partial x_j=\partial/\partial r_j$ inside the integral, so we can apply Gauss's theorem in the $\mathbf{r}-$space, obtain
\begin{equation}
    \widetilde A_{ij}
    =
    \frac{1}{V_\ell}
    \oint_{S_\ell}
    u_i(\mathbf{x}+\mathbf r)\hat n_j\,dS ,
\label{pos_r}
\end{equation}
where $S_\ell=4\pi r^2$ and
$\hat{\mathbf n}=\mathbf r/r$ is the outward unit normal vector. The same
filtered velocity can equivalently be written using
$u_i(\mathbf{x}-\mathbf r)$, which gives
\begin{equation}
    \widetilde A_{ij}
    =
    -\frac{1}{V_\ell}
    \oint_{S_\ell}
    u_i(\mathbf{x}-\mathbf r)\hat n_j\,dS .
\label{neg_r}
\end{equation}
Adding Eqs.~\eqref{pos_r} and \eqref{neg_r} yields
\begin{equation}
    2\widetilde A_{ij}
    =
    \frac{1}{V_\ell}
    \oint_{S_\ell}
    \delta u_i\,\hat n_j\,dS .
\end{equation}
Since $V_\ell/S_\ell=r/3=\ell/6$, this can be written as the exact
identity
\begin{equation}
    \ell\widetilde A_{ij}
    =
    \frac{3}{S_\ell}
    \oint_{S_\ell}
    \delta u_i\,\hat n_j\,dS .
\label{du_VGT_relation}
\end{equation}

We note that Eq.~\eqref{du_VGT_relation} has a natural interpretation in terms
of Spherical Harmonics. At fixed $\mathbf{x}$ and fixed radius $r=\ell/2$, the velocity increment
$\delta u_i$ depends only on the normal direction
$\hat{\mathbf n}$. Therefore we could think of each component $i$ as an angular field defined on the sphere, and then apply the spherical-harmonic
decomposition
\begin{equation}
    \delta u_i(\hat{\mathbf n})
    =
    \sum_{L=0}^{\infty}
    \delta u_i^{(L)}(\hat{\mathbf n}),
\end{equation}
where superscript $L$ denotes the spherical-harmonic degree. $L=1$ (usually referred to as the dipolar sector) is three-dimensional and
can be represented by the Cartesian coordinates
$\{\hat n_1,\hat n_2,\hat n_3\}$. Therefore, for
each velocity component $i$, the projection of the angular field
$\delta u_i(\hat{\mathbf n})$ onto this subspace can be written as
\begin{equation}
    \delta u_i^{(1)}(\hat{\mathbf n})
    =
    B_{ij}\hat n_j ,
\label{dipole_form}
\end{equation}
where $B_{ij}$ are the projection coefficients to be determined. 

To obtain $B_{ij}$, we can multiply Eq.~\eqref{dipole_form} by
$\hat n_k$ and average over the spherical surface:
\begin{equation}
    \frac{1}{S_\ell}
    \oint_{S_\ell}
    \delta u_i^{(1)}\hat n_k\,dS
    =
    B_{ij}
    \frac{1}{S_\ell}
    \oint_{S_\ell}
    \hat n_j\hat n_k\,dS .
\end{equation}
Spherical symmetry gives the exact second-order angular moment
\begin{equation}
    \frac{1}{S_\ell}
    \oint_{S_\ell}
    \hat n_j\hat n_k\,dS
    =
    \frac{1}{3}\delta_{jk},
\end{equation}
where $\delta_{jk}$ is the Kronecker delta. Hence,
\begin{equation}
    \frac{1}{S_\ell}
    \oint_{S_\ell}
    \delta u_i^{(1)}\hat n_k\,dS
    =
    \frac{1}{3}B_{ik}.
\end{equation}
Since $\delta u_i^{(1)}$ is the orthogonal projection of the full
field $\delta u_i$ onto the $L=1$ subspace, the higher-order
spherical-harmonic components are orthogonal to $\hat n_k$. Therefore,
\begin{equation}
    \frac{1}{S_\ell}
    \oint_{S_\ell}
    \delta u_i^{(1)}\hat n_k\,dS
    =
    \frac{1}{S_\ell}
    \oint_{S_\ell}
    \delta u_i\hat n_k\,dS .
\end{equation}
Combining the previous two relations gives
\begin{equation}
    B_{ik}
    =
    \frac{3}{S_\ell}
    \oint_{S_\ell}
    \delta u_i\hat n_k\,dS .
\end{equation} Remarkably, it is exactly equal to the right-hand side of Eq.~\eqref{du_VGT_relation}. Thus,
\begin{equation}
    B_{ij}
    =
    \ell\widetilde A_{ij}.
\end{equation}
Hence the dipolar component of the finite velocity increment is
determined exactly by the filtered velocity-gradient tensor,
\begin{equation}
    \delta u_i^{(1)}(\hat{\mathbf n})
    =
    \ell\widetilde A_{ij}\hat n_j
    =
    2\widetilde A_{ij}r_j .
\label{dipole_increment}
\end{equation}

The symmetric velocity increment is odd under reversal of the
separation direction, i.e.,
$
\delta u_i(-\hat{\mathbf n})
=
-\delta u_i(\hat{\mathbf n}).
$
Consequently, all even spherical-harmonic degrees vanish, including the monopole $L=0$ component. The exact angular decomposition can
therefore be written as
\begin{equation}
    \delta u_i
    =
    2\widetilde A_{ij}r_j
    +
    q_i(\hat{\mathbf n}),
\label{du_est}
\end{equation}
where
\begin{equation}
    q_i(\hat{\mathbf n})
    =
    \sum_{L=3,5,7,\ldots}^{\infty}
    \delta u_i^{(L)}(\hat{\mathbf n})
\end{equation}
contains all higher odd angular modes. By construction, this residual
is orthogonal to the dipolar subspace, $
    \frac{1}{S_\ell}
    \oint_{S_\ell}
    q_i(\hat{\mathbf n})\hat n_j\,dS
    =
    0 .$ Equation~\eqref{du_est} therefore has a clear physical meaning. The term
$2\widetilde A_{ij}r_j$ is the exact $L=1$ spherical-harmonic projection of the finite velocity-increment field on the bounding
sphere. The residual $q_i$ contains the higher-order angular structure that cannot be represented by the rank-two coarse-grained
velocity-gradient tensor (VGT).

\section{Eigenvalue Decomposition and Transformation}

In this section, we provide the details leading to the VGT representation of the scale-space energy flux
\begin{equation}
    \Phi_{\ell}(\mathbf{x},t)
    \equiv
    -\frac{3}{4\ell}\frac{1}{S_{\ell}}
    \oint_{S_{\ell}}
    \delta u_i^2\,\delta u_j\hat{n}_j\,dS,
\label{cascade}
\end{equation}
where $S_\ell=4\pi r^2$ is the surface area of the sphere of radius
$r=\ell/2$.

Using the estimation of $\delta u_i$ derived in the preceding section,
$\delta u_i\approx2\widetilde A_{ij}r_j$, the squared velocity
increment becomes
\begin{equation}
    \delta u_i^2
    \approx
    4
    \|\widetilde{\mathbf A}\mathbf r\|^2
    =
    4\
    (\widetilde{\mathbf A}\mathbf r)^T
    (\widetilde{\mathbf A}\mathbf r),
\label{du2_decom}
\end{equation}
where
$\mathbf r=[r_1,r_2,r_3]^T$ and $r=\|\mathbf r\|$.

The longitudinal velocity increment can similarly be written as
\begin{align}
    \delta u_j\hat n_j
    &\approx
    2\widetilde A_{jk}r_k\frac{r_j}{r}
\nonumber\\
    &=
    \frac{2}{r}
    \mathbf r^T\widetilde{\mathbf A}\mathbf r
\nonumber\\
    &=
    \frac{2}{r}
    \mathbf r^T
    (\widetilde{\mathbf S}+\widetilde{\boldsymbol\Omega})
    \mathbf r
\nonumber\\
    &=
    \frac{2}{r}
    \mathbf r^T\widetilde{\mathbf S}\mathbf r ,
\label{longi_matrix}
\end{align}
where
\begin{equation}
    \widetilde S_{ij}
    =
    \frac{1}{2}
    (\widetilde A_{ij}+\widetilde A_{ji}),
    \qquad
    \widetilde\Omega_{ij}
    =
    \frac{1}{2}
    (\widetilde A_{ij}-\widetilde A_{ji})
\end{equation}
are the filtered strain-rate and rotation-rate tensors, respectively.
The last equality in Eq.~\eqref{longi_matrix} follows from the
antisymmetry of $\widetilde{\boldsymbol\Omega}$, for which
$\mathbf r^T\widetilde{\boldsymbol\Omega}\mathbf r=0$. Because $\widetilde{\mathbf S}$ is real and symmetric, it admits the
eigendecomposition
\begin{equation}
    \widetilde{\mathbf S}
    =
    \mathbf X
    \widetilde{\boldsymbol\Lambda}^{S}
    \mathbf X^T,
\end{equation}
where
$\mathbf X=[\mathbf X_1,\mathbf X_2,\mathbf X_3]$ contains the
orthonormal strain eigenvectors and
\begin{equation}
    \widetilde{\boldsymbol\Lambda}^{S}
    =
    {\rm diag}
    (\widetilde\lambda_1,
     \widetilde\lambda_2,
     \widetilde\lambda_3)
\end{equation}
contains the corresponding eigenvalues. For incompressible flow,
\begin{equation}
    \widetilde\lambda_1+
    \widetilde\lambda_2+
    \widetilde\lambda_3=0.
\end{equation}
The eigenvectors are chosen to form a right-handed orthonormal basis,
so that $\mathbf X^T\mathbf X=\mathbf X\mathbf X^T=\mathbf I$ and
$\det\mathbf X=1$. Introducing the coordinates in the strain eigenbasis,
\begin{equation}
    \mathbf r^*
    \equiv
    \mathbf X^T\mathbf r,
\end{equation}
the longitudinal increment becomes
\begin{align}
    \delta u_j\hat n_j
    &\approx
    \frac{2}{r}
    \mathbf r^T
    \mathbf X\widetilde{\boldsymbol\Lambda}^{S}
    \mathbf X^T\mathbf r
\nonumber\\
    &=
    \frac{2}{r}
    \mathbf r^{*T}
    \widetilde{\boldsymbol\Lambda}^{S}
    \mathbf r^*.
\label{longi_decom}
\end{align}

Substituting Eqs.~\eqref{du2_decom} and \eqref{longi_decom} into
Eq.~\eqref{cascade} gives
\begin{equation}
\Phi_{\ell}(\mathbf{x},t)
\approx
-\frac{3}{4\ell r}
\frac{8}{S_{\ell}}
\oint_{S_{\ell}}
(\widetilde{\mathbf A}\mathbf r)^T
(\widetilde{\mathbf A}\mathbf r)
\,
\mathbf r^{*T}
\widetilde{\boldsymbol\Lambda}^{S}
\mathbf r^*
\,dS .
\label{4.19}
\end{equation}

To express the first quadratic form in the same coordinates, define
\begin{equation}
    \widetilde{\boldsymbol\Omega}^{*}
    \equiv
    \mathbf X^T
    \widetilde{\boldsymbol\Omega}
    \mathbf X.
\end{equation}
Then
\begin{equation}
    \mathbf X^T\widetilde{\mathbf A}\mathbf X
    =
    \widetilde{\boldsymbol\Lambda}^{S}
    +
    \widetilde{\boldsymbol\Omega}^{*},
\end{equation}
and hence
\begin{align}
    (\widetilde{\mathbf A}\mathbf r)^T
    (\widetilde{\mathbf A}\mathbf r)
    &=
    \left[
    (\widetilde{\boldsymbol\Lambda}^{S}
    +\widetilde{\boldsymbol\Omega}^{*})
    \mathbf r^*
    \right]^T
    \left[
    (\widetilde{\boldsymbol\Lambda}^{S}
    +\widetilde{\boldsymbol\Omega}^{*})
    \mathbf r^*
    \right]
\nonumber\\
    &=
    \left\|
    (\widetilde{\boldsymbol\Lambda}^{S}
    +\widetilde{\boldsymbol\Omega}^{*})
    \mathbf r^*
    \right\|^2.
\label{4.21}
\end{align}
Because the transformation $\mathbf r\mapsto\mathbf r^*$ is
orthogonal, both the spherical surface and its area element $dS$
are invariant. Equation~\eqref{4.19} can therefore be evaluated
directly in the strain eigenbasis. Writing
\begin{equation}
    \mathbf r^*
    =
    \begin{bmatrix}
    r_1\\r_2\\r_3
    \end{bmatrix},
    \qquad
    \widetilde{\boldsymbol\Lambda}^{S}
    =
    \begin{bmatrix}
    \widetilde\lambda_1&0&0\\
    0&\widetilde\lambda_2&0\\
    0&0&\widetilde\lambda_3
    \end{bmatrix},
\end{equation}
and
\begin{equation}
    \widetilde{\boldsymbol\Omega}^{*}
    =
    \begin{bmatrix}
    0&
    \widetilde\Omega_{12}^{*}&
    \widetilde\Omega_{13}^{*}\\
    -\widetilde\Omega_{12}^{*}&
    0&
    \widetilde\Omega_{23}^{*}\\
    -\widetilde\Omega_{13}^{*}&
    -\widetilde\Omega_{23}^{*}&
    0
    \end{bmatrix},
\end{equation}
we have
\begin{equation}
(\widetilde{\boldsymbol\Lambda}^{S}
+\widetilde{\boldsymbol\Omega}^{*})
\mathbf r^*
=
\begin{bmatrix}
\widetilde\lambda_1r_1
+\widetilde\Omega_{12}^{*}r_2
+\widetilde\Omega_{13}^{*}r_3
\\
-\widetilde\Omega_{12}^{*}r_1
+\widetilde\lambda_2r_2
+\widetilde\Omega_{23}^{*}r_3
\\
-\widetilde\Omega_{13}^{*}r_1
-\widetilde\Omega_{23}^{*}r_2
+\widetilde\lambda_3r_3
\end{bmatrix}.
\end{equation}

The squared norm contains terms proportional to $r_i^2$ and cross
terms proportional to $r_ir_j$. The diagonal terms are
\begin{equation}
\begin{aligned}
&
\widetilde\lambda_1^2r_1^2
+\widetilde\Omega_{12}^{*2}r_2^2
+\widetilde\Omega_{13}^{*2}r_3^2
+\widetilde\Omega_{12}^{*2}r_1^2
+\widetilde\lambda_2^2r_2^2
+\widetilde\Omega_{23}^{*2}r_3^2
\\
&\qquad
+\widetilde\Omega_{13}^{*2}r_1^2
+\widetilde\Omega_{23}^{*2}r_2^2
+\widetilde\lambda_3^2r_3^2 ,
\end{aligned}
\end{equation}
whereas the cross terms are
\begin{equation}
\begin{aligned}
&
2\left[
\widetilde\Omega_{12}^{*}
(\widetilde\lambda_1-\widetilde\lambda_2)
+
\widetilde\Omega_{13}^{*}\widetilde\Omega_{23}^{*}
\right]r_1r_2
\\
&+
2\left[
\widetilde\Omega_{13}^{*}
(\widetilde\lambda_1-\widetilde\lambda_3)
-
\widetilde\Omega_{12}^{*}\widetilde\Omega_{23}^{*}
\right]r_1r_3
\\
&+
2\left[
\widetilde\Omega_{12}^{*}\widetilde\Omega_{13}^{*}
+
\widetilde\Omega_{23}^{*}
(\widetilde\lambda_2-\widetilde\lambda_3)
\right]r_2r_3 .
\end{aligned}
\end{equation}

In Eq.~\eqref{4.19}, these cross terms are multiplied by
\begin{equation}
    \mathbf r^{*T}
    \widetilde{\boldsymbol\Lambda}^{S}
    \mathbf r^*
    =
    \widetilde\lambda_1r_1^2
    +
    \widetilde\lambda_2r_2^2
    +
    \widetilde\lambda_3r_3^2.
\end{equation}
Every resulting cross contribution therefore contains an odd power
of at least one Cartesian coordinate. Its integral over the
spherically symmetric surface vanishes by parity. For example,
reflection $r_1\rightarrow-r_1$ leaves the integration surface
unchanged while reversing the sign of any term containing an odd
power of $r_1$.

Only even fourth-order moments consequently remain, and
Eq.~\eqref{4.19} reduces to
\begin{equation}
\begin{aligned}
\Phi_{\ell}(\mathbf{x},t)
\approx
-\frac{3}{4\ell r}\frac{8}{S_{\ell}}
\oint_{S_{\ell}}
\big[
&\widetilde{\lambda}_1^2r_1^2
+\widetilde{\Omega}_{12}^{*2}r_2^2
+\widetilde{\Omega}_{13}^{*2}r_3^2
\\
&+\widetilde{\lambda}_2^2r_2^2
+\widetilde{\Omega}_{12}^{*2}r_1^2
+\widetilde{\Omega}_{23}^{*2}r_3^2
\\
&+\widetilde{\lambda}_3^2r_3^2
+\widetilde{\Omega}_{13}^{*2}r_1^2
+\widetilde{\Omega}_{23}^{*2}r_2^2
\big]
\\
&\times
(\widetilde{\lambda}_1r_1^2
+\widetilde{\lambda}_2r_2^2
+\widetilde{\lambda}_3r_3^2)\,dS .
\end{aligned}
\end{equation}

By spherical symmetry, only two distinct fourth-order surface
integrals are required,
\begin{equation}
    A
    \equiv
    \oint_{S_\ell}r_i^4\,dS
    =
    \frac{4\pi}{5}r^6 ,
\end{equation}
and
\begin{equation}
    B
    \equiv
    \oint_{S_\ell}r_i^2r_j^2\,dS
    =
    \frac{4\pi}{15}r^6,
    \qquad i\neq j,
\end{equation}
so that $A=3B$. Using this relation together with incompressibility,
$\widetilde\lambda_1+\widetilde\lambda_2+\widetilde\lambda_3=0$,
gives
\begin{equation}
\Phi_{\ell}(\mathbf{x},t)
\approx
-\frac{12B}{\ell rS_{\ell}}
\left(
\widetilde\lambda_1^3
+\widetilde\lambda_2^3
+\widetilde\lambda_3^3
\\
-\widetilde\Omega_{23}^{*2}\widetilde\lambda_1
-\widetilde\Omega_{13}^{*2}\widetilde\lambda_2
-\widetilde\Omega_{12}^{*2}\widetilde\lambda_3
\right).
\end{equation}
With $B=4\pi r^6/15$, this becomes
\begin{equation}
\Phi_{\ell}(\mathbf{x},t)
\approx
-\frac{2}{5}r^2
\left(
\widetilde\lambda_1^3
+\widetilde\lambda_2^3
+\widetilde\lambda_3^3
-\widetilde\Omega_{23}^{*2}\widetilde\lambda_1
-\widetilde\Omega_{13}^{*2}\widetilde\lambda_2
-\widetilde\Omega_{12}^{*2}\widetilde\lambda_3
\right).
\label{Phi_cascade}
\end{equation}

The strain self-amplification (SSA) invariant is defined as
\begin{equation}
    \widetilde R_s
    \equiv
    -\frac{1}{3}
    \widetilde S_{ij}
    \widetilde S_{jk}
    \widetilde S_{ki}.
    \label{SSA}
\end{equation}
In the strain eigenbasis,
\begin{equation}
    \widetilde R_s
    =
    -\frac{1}{3}
    \left(
    \widetilde\lambda_1^3+
    \widetilde\lambda_2^3+
    \widetilde\lambda_3^3
    \right)
    =
    -\widetilde\lambda_1
     \widetilde\lambda_2
     \widetilde\lambda_3,
\end{equation}
where the latter equality follows from
$\widetilde\lambda_1+\widetilde\lambda_2+\widetilde\lambda_3=0$.

The filtered rotation tensor is related to the filtered vorticity
$\widetilde{\boldsymbol\omega}
=\nabla\times\widetilde{\mathbf u}$ according to
\begin{equation}
    \widetilde\Omega_{ij}
    =
    -\frac{1}{2}
    \epsilon_{ijk}\widetilde\omega_k.
\end{equation}
Because the strain eigenbasis is chosen to be right handed,
\begin{equation}
    \widetilde{\boldsymbol\omega}^{*}
    =
    \mathbf X^T
    \widetilde{\boldsymbol\omega},
\end{equation}
and hence
\begin{equation}
    \widetilde\Omega_{23}^{*2}
    =
    \frac14\widetilde\omega_1^{*2},
    \qquad
    \widetilde\Omega_{13}^{*2}
    =
    \frac14\widetilde\omega_2^{*2},
    \qquad
    \widetilde\Omega_{12}^{*2}
    =
    \frac14\widetilde\omega_3^{*2}.
\end{equation}
The vortex-stretching (VS) invariant is
\begin{equation}
    \widetilde V_s
    \equiv
    \widetilde\omega_i
    \widetilde S_{ij}
    \widetilde\omega_j
    =
    \widetilde{\boldsymbol\omega}^{*T}
    \widetilde{\boldsymbol\Lambda}^{S}
    \widetilde{\boldsymbol\omega}^{*}
    =
    \sum_{i=1}^{3}
    \widetilde\lambda_i
    \widetilde\omega_i^{*2}.
    \label{VS}
\end{equation}

Substituting Eq. \ref{SSA} and \ref{VS} into Eq. \ref{Phi_cascade}, the cascade rate therefore becomes
\begin{equation}
    \Phi_{\ell}(\mathbf{x},t)
    \approx
    \frac{2}{5}r^2
    \left(
    3\widetilde R_s
    +
    \frac14\widetilde V_s
    \right).
\end{equation}

It is useful to compare this result with the corresponding
filtered-space energy flux under top-hat filter. Following Borue and Orszag
\cite{borue1998local}, its leading  representation can be written
as
\begin{equation}
\Pi_{\ell}
\approx
\frac{3}
{11}r^2
\left(
3\widetilde R_s+\frac14\widetilde V_s
\right).
\end{equation}

\section{Longitudinal and Transverse Orthogonal Decomposition}

In this section, we provide the detailed derivation of the longitudinal--transverse orthogonal decomposition used in the main text. For a unit vector
$\hat{\mathbf n}=\mathbf r/r$ normal to the spherical surface, the
velocity increment can be decomposed into longitudinal and transverse
parts, as
\begin{equation}
    |\delta\mathbf u|^2
    =
    (\delta\mathbf u\cdot\hat{\mathbf n})^2
    +
    \|\delta\mathbf u\times\hat{\mathbf n}\|^2,
\end{equation}
the local scale-space flux can be then decomposed exactly as
\begin{align}
\Phi_{\ell}(\mathbf{x},t)
&\equiv
-\frac{3}{4\ell}\frac{1}{S_{\ell}}
\oint_{S_{\ell}}
(\delta\mathbf{u}\cdot\hat{\mathbf{n}})^3\,dS
\nonumber\\
&\quad-
\frac{3}{4\ell}\frac{1}{S_{\ell}}
\oint_{S_{\ell}}
(\delta\mathbf{u}\cdot\hat{\mathbf{n}})
\left\|
\delta\mathbf{u}\times\hat{\mathbf{n}}
\right\|^2
\,dS
\nonumber\\
&\equiv
\Phi_{\ell}^{\parallel}
+
\Phi_{\ell}^{\perp}.
\label{LTOD_supp}
\end{align}
Here $\Phi_\ell^\parallel$ and $\Phi_\ell^\perp$ denote the
longitudinal and transverse contributions, respectively.

We first consider the longitudinal contribution. Using the estimation of $\delta u_i$ result
derived in the preceding section,
$\delta u_i\approx2\widetilde A_{ij}r_j$, gives
\begin{equation}
\Phi_{\ell}^{\parallel}
=
-\frac{3}{4\ell}\frac{1}{S_{\ell}}
\oint_{S_{\ell}}
(\delta\mathbf{u}\cdot\hat{\mathbf{n}})^3\,dS
\approx
-\frac{3}{4\ell}\frac{1}{S_{\ell}}
\oint_{S_{\ell}}
\left(
\frac{2}{r}
\mathbf r^{T}\widetilde{\mathbf A}\mathbf r
\right)^3
\,dS .
\end{equation}
Because the antisymmetric rotation tensor satisfies
$\mathbf r^T\widetilde{\boldsymbol\Omega}\mathbf r=0$,
only the strain-rate tensor contributes to the longitudinal
increment. In the eigenbasis of $\widetilde{\mathbf S}$,
introduced in the preceding section,
\begin{equation}
    \mathbf r^T\widetilde{\mathbf A}\mathbf r
    =
    \mathbf r^T\widetilde{\mathbf S}\mathbf r
    =
    \widetilde\lambda_1r_1^2
    +
    \widetilde\lambda_2r_2^2
    +
    \widetilde\lambda_3r_3^2 ,
\end{equation}
where $\widetilde\lambda_i$ are the three eigenvalues of the filtered
strain-rate tensor. Hence
\begin{align}
\Phi_{\ell}^{\parallel}
&\approx
-\frac{3}{4\ell}
\frac{8}{r^3S_{\ell}}
\oint_{S_{\ell}}
\left(
\widetilde{\lambda}_1r_1^2
+
\widetilde{\lambda}_2r_2^2
+
\widetilde{\lambda}_3r_3^2
\right)^3
\,dS
\nonumber\\
&=
-\frac{3}{4\ell}
\frac{8}{r^3S_{\ell}}
\left(
8\frac{4\pi}{35}r^8
\widetilde{\lambda}_1
\widetilde{\lambda}_2
\widetilde{\lambda}_3
\right).
\end{align}
The second equality follows from the isotropic sixth-order moments on
a sphere together with incompressibility,
$\widetilde\lambda_1+\widetilde\lambda_2+\widetilde\lambda_3=0$.
Using
\begin{equation}
    \widetilde R_s
    =
    -\widetilde\lambda_1
     \widetilde\lambda_2
     \widetilde\lambda_3,
\end{equation} we obtain
\begin{equation}
    \Phi_{\ell}^{\parallel}
    \approx
    \frac{24}{35}
    r^2\widetilde R_s .
\label{Phi_parallel_supp}
\end{equation}

We next consider the transverse contribution,
\begin{equation}
\Phi_{\ell}^{\perp}
\approx
-\frac{3}{4\ell}\frac{1}{S_{\ell}}
\oint_{S_{\ell}}
\frac{2}{r}
\mathbf r^{T}\widetilde{\mathbf A}\mathbf r
\left\|
\delta\mathbf{u}\times\hat{\mathbf{n}}
\right\|^2
\,dS .
\end{equation}
Using
$\delta\mathbf u\approx2\widetilde{\mathbf A}\mathbf r$
and the identity
$|\mathbf a\times\hat{\mathbf n}|^2
=|\mathbf a|^2-(\mathbf a\cdot\hat{\mathbf n})^2$,
we obtain
\begin{equation}
\left\|
\delta\mathbf{u}\times\hat{\mathbf{n}}
\right\|^2
\approx
\frac{4}{r^2}
\left[
r^2
\left\|
\widetilde{\mathbf A}\mathbf r
\right\|^2
-
\left(
\mathbf r^{T}\widetilde{\mathbf A}\mathbf r
\right)^2
\right].
\label{transverse_increment}
\end{equation}
Substituting Eq.~\eqref{transverse_increment} into the transverse
flux gives
\begin{equation}
\begin{aligned}
\Phi_{\ell}^{\perp}
&\approx
-\frac{3}{4\ell}
\frac{1}{S_{\ell}}
\oint_{S_{\ell}}
\frac{2}{r}
\mathbf r^{T}\widetilde{\mathbf A}\mathbf r
\left[
4\left\|
\widetilde{\mathbf A}\mathbf r
\right\|^2
-
\frac{4}{r^2}
\left(
\mathbf r^{T}\widetilde{\mathbf A}\mathbf r
\right)^2
\right]
\,dS
\\
&=
\frac{2}{5}r^2
\left(
3\widetilde R_s+\frac{1}{4}\widetilde V_s
\right)
-
\frac{24}{35}r^2\widetilde R_s .
\end{aligned}
\end{equation}
In the second equality, the first contribution is the $L=1$ VGT-resolved
expression for the total flux derived in the preceding section,
whereas the second is the longitudinal contribution
Eq.~\eqref{Phi_parallel_supp}. After simplification,
\begin{equation}
    \Phi_{\ell}^{\perp}
    \approx
    \frac{18}{35}
    r^2\widetilde R_s
    +
    \frac{1}{10}
    r^2\widetilde V_s .
\label{Phi_perp_supp}
\end{equation}

\section{Conditional statistics at additional length scales}

To examine the robustness of the conditional statistics across the dissipation and
inertial ranges, we supplement the results at $\ell=45\eta$ shown in
the main text with joint conditional averages at
$\ell/\eta=15$, $30$, $60$, and $75$. Throughout this section,
$\Phi_\ell$, $\Phi_\ell^{\parallel}$, and $\Phi_\ell^{\perp}$ are
normalized by the mean dissipation rate $\langle\epsilon\rangle$,
while $\widetilde R_s$ and $\widetilde V_s$ are normalized by
$
[\,\frac{1}{2}
\langle\widetilde\Omega_{ij}\widetilde\Omega_{ij}\rangle\,]^{3/2},
$
where
$\widetilde\Omega_{ij}
=(\partial_j\widetilde u_i-\partial_i\widetilde u_j)/2$
is the filtered rotation-rate tensor.

\begin{figure}[htbp]
    \centering
    \includegraphics[width=0.4\textwidth]{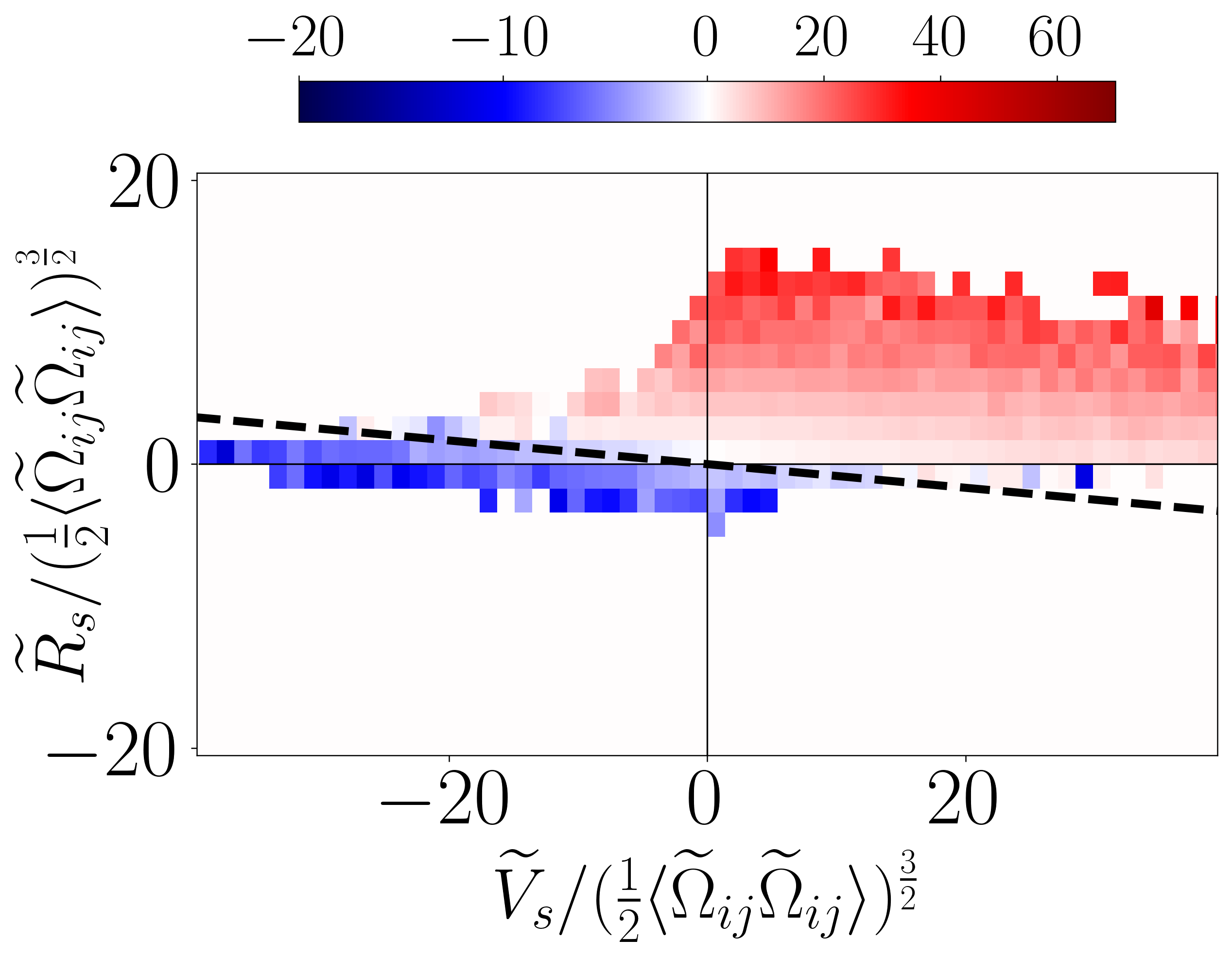}
    \caption{Joint conditional average
    $\langle\Phi_\ell\mid\widetilde R_s,\widetilde V_s\rangle/
    \langle\epsilon\rangle$
    at $\ell=15\eta$. Red and blue regions denote positive
    (forward) and negative (inverse) conditional mean energy transfer,
    respectively. The dashed line represents the analytical prediction of forward and inverse cascade, boundary $3\widetilde{R}_s+\widetilde{V}_s/4=0$.}
    \label{Con_stat_15}
\end{figure}

\begin{figure}[htbp]
    \centering
    \includegraphics[width=0.4\textwidth]{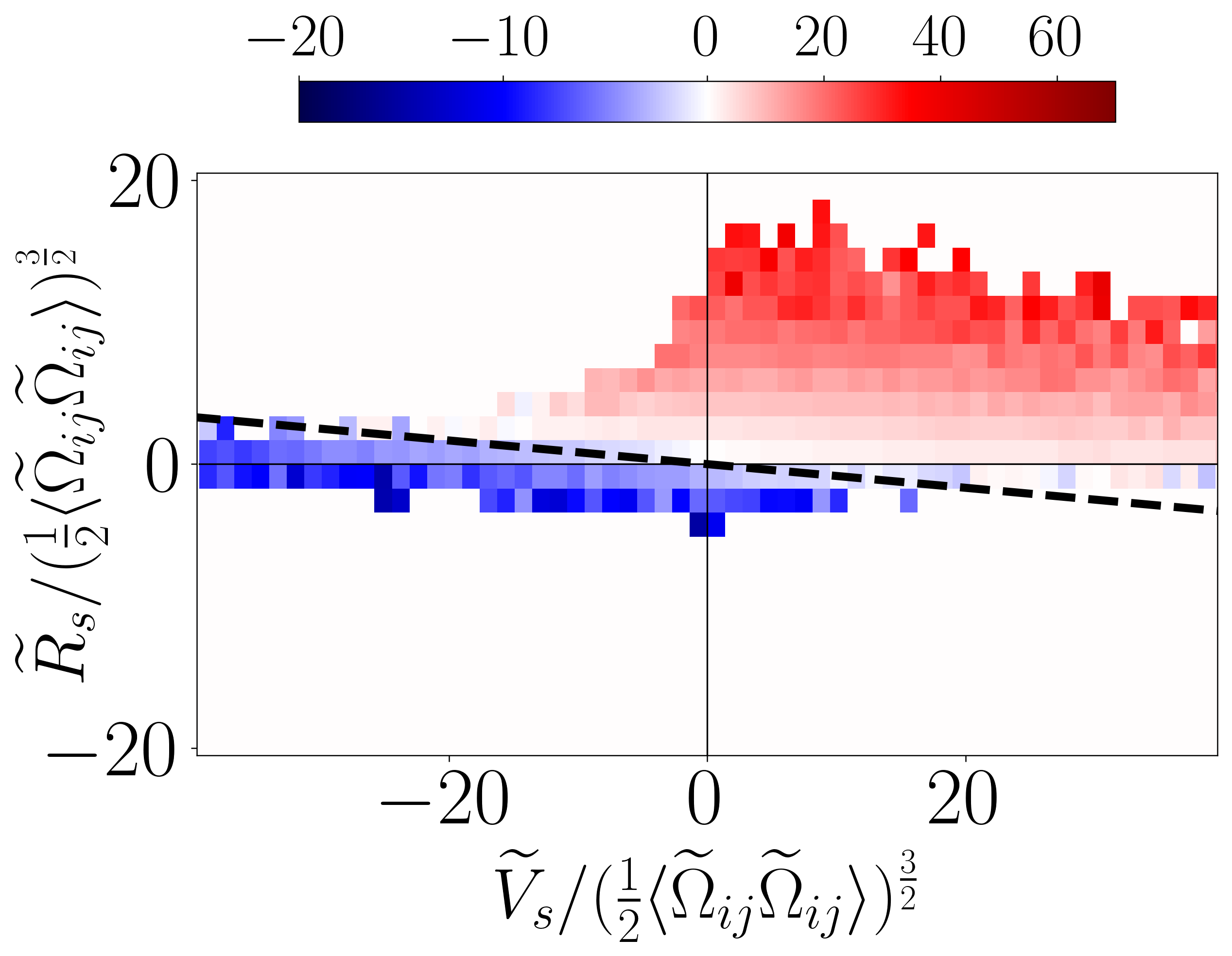}
    \caption{Same as Fig.~\ref{Con_stat_15}, but for
    $\ell=30\eta$.}
    \label{Con_stat_30}
\end{figure}

\begin{figure}[htbp]
    \centering
    \includegraphics[width=0.4\textwidth]{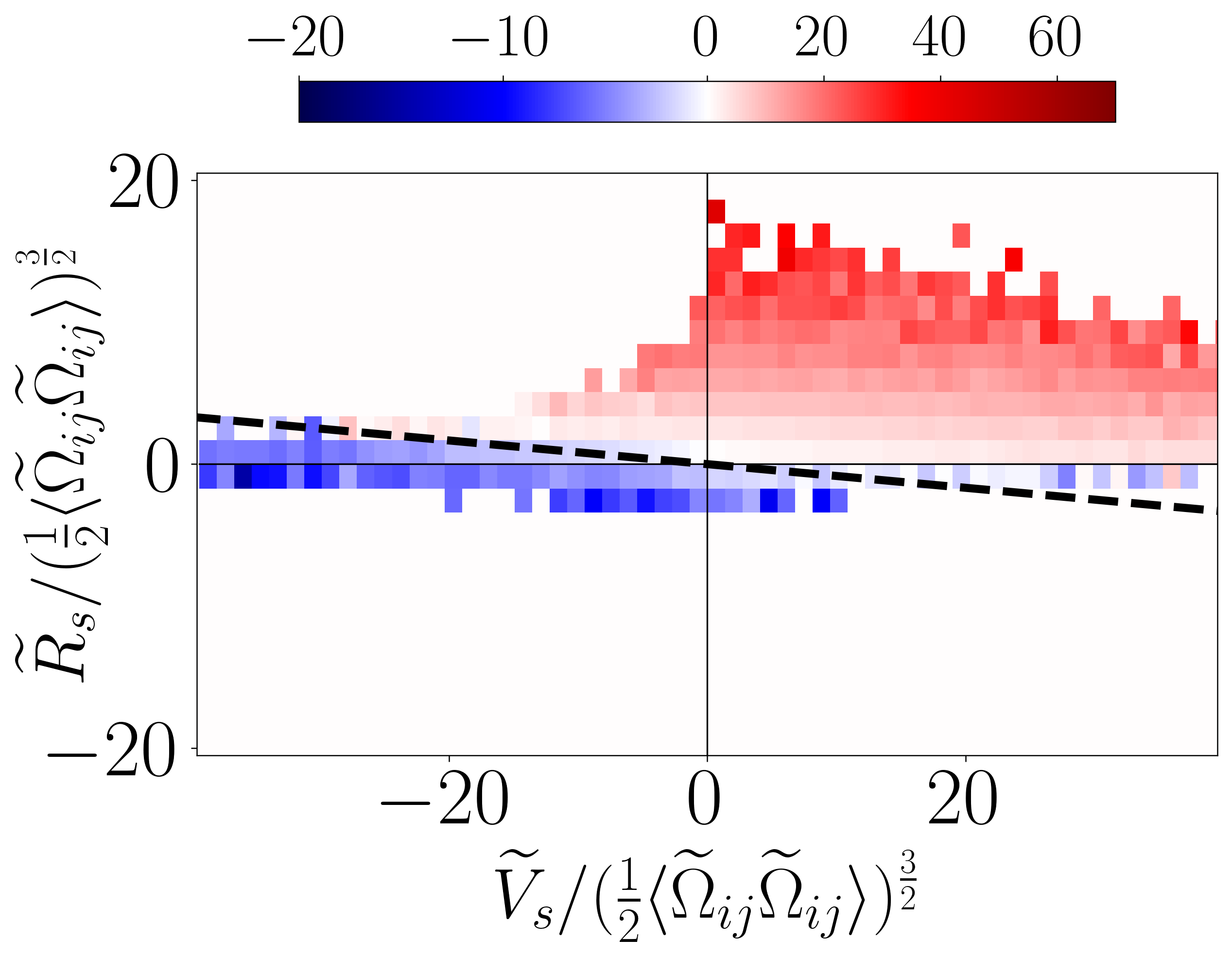}
    \caption{Same as Fig.~\ref{Con_stat_15}, but for
    $\ell=60\eta$.}
    \label{Con_stat_60}
\end{figure}

\begin{figure}[htbp]
    \centering
    \includegraphics[width=0.4\textwidth]{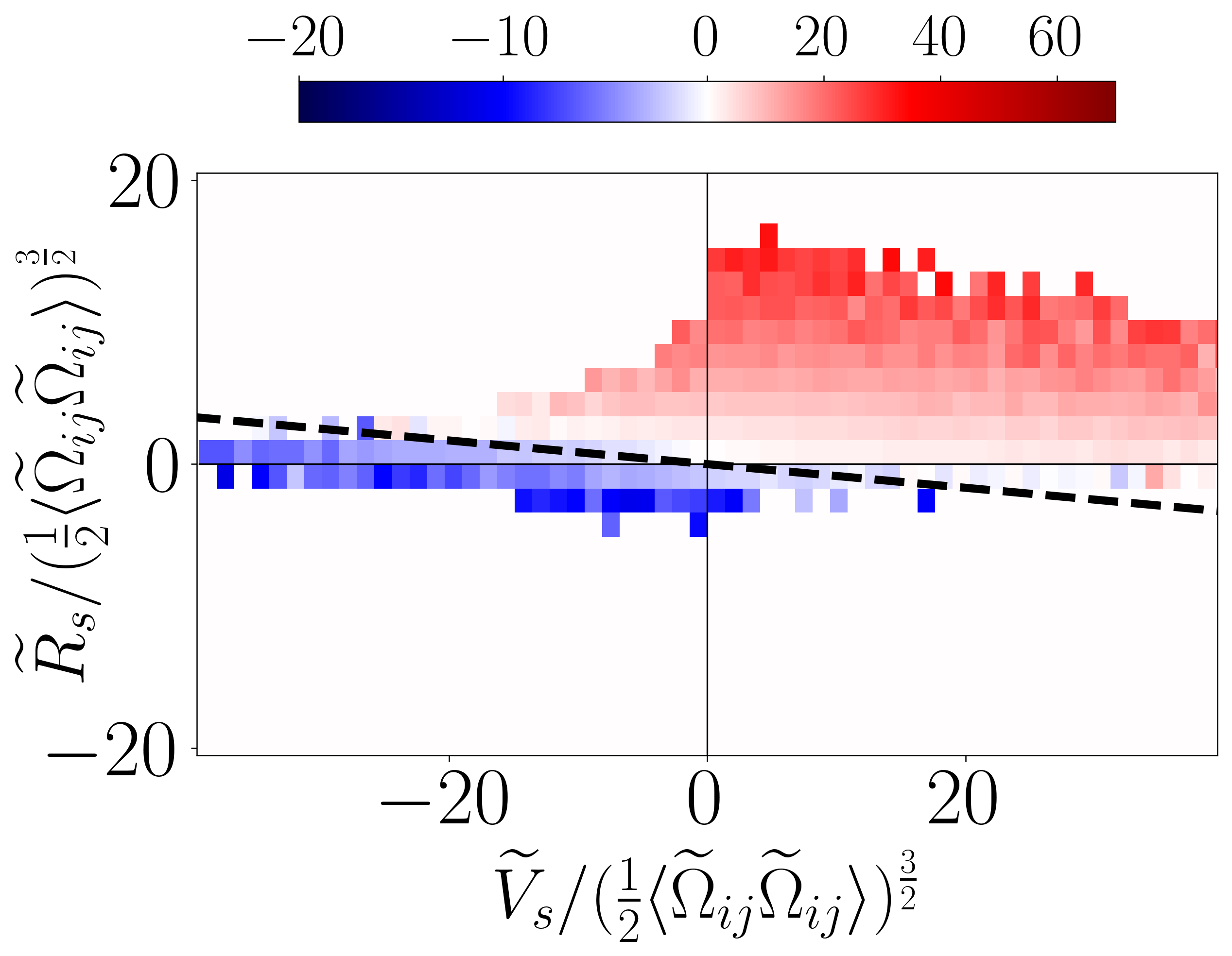}
    \caption{Same as Fig.~\ref{Con_stat_15}, but for
    $\ell=75\eta$.}
    \label{Con_stat_75}
\end{figure}

\begin{figure}[htbp]
    \centering
    \includegraphics[width=0.235\textwidth]{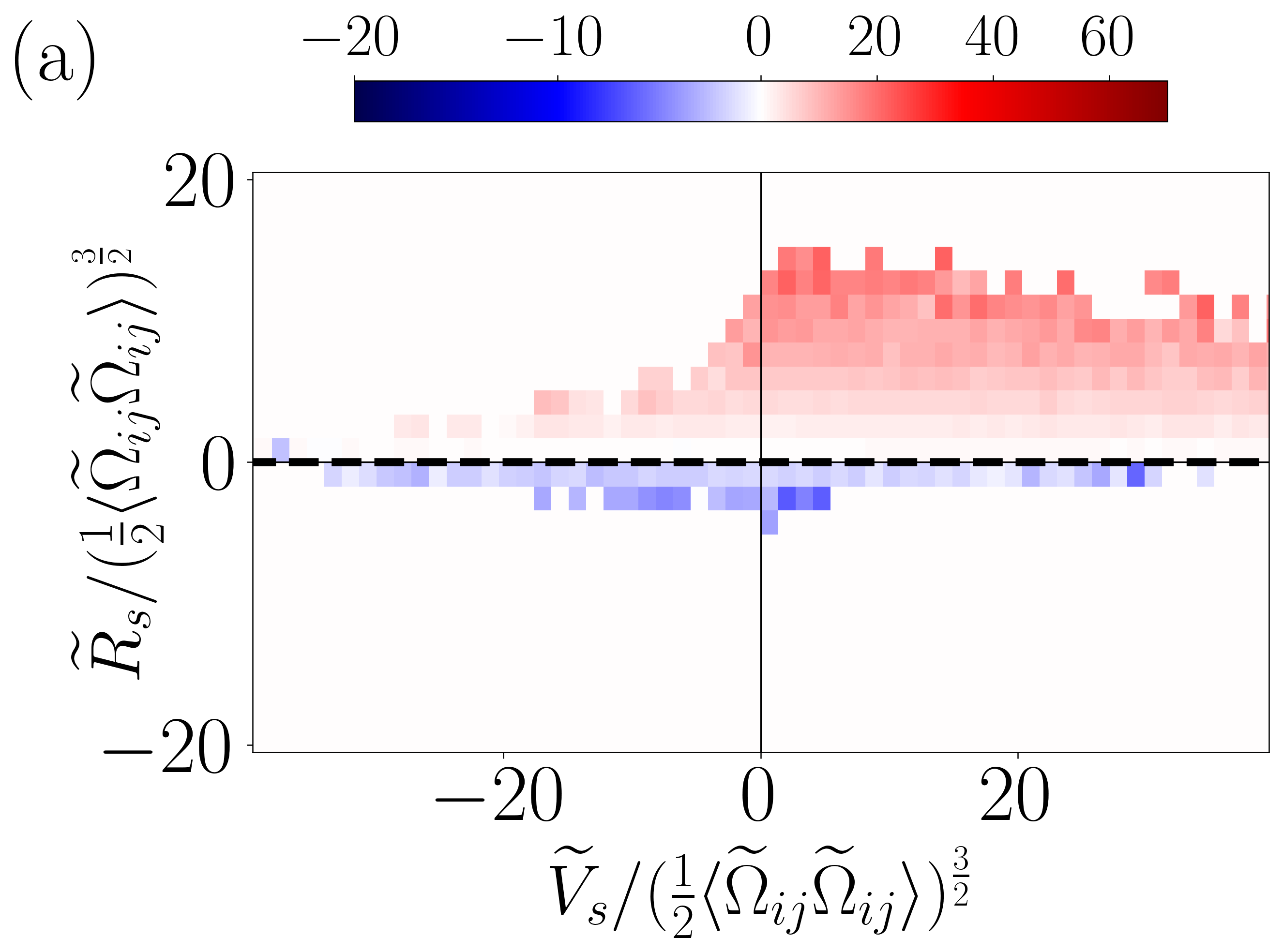}
    \hfill
    \includegraphics[width=0.235\textwidth]{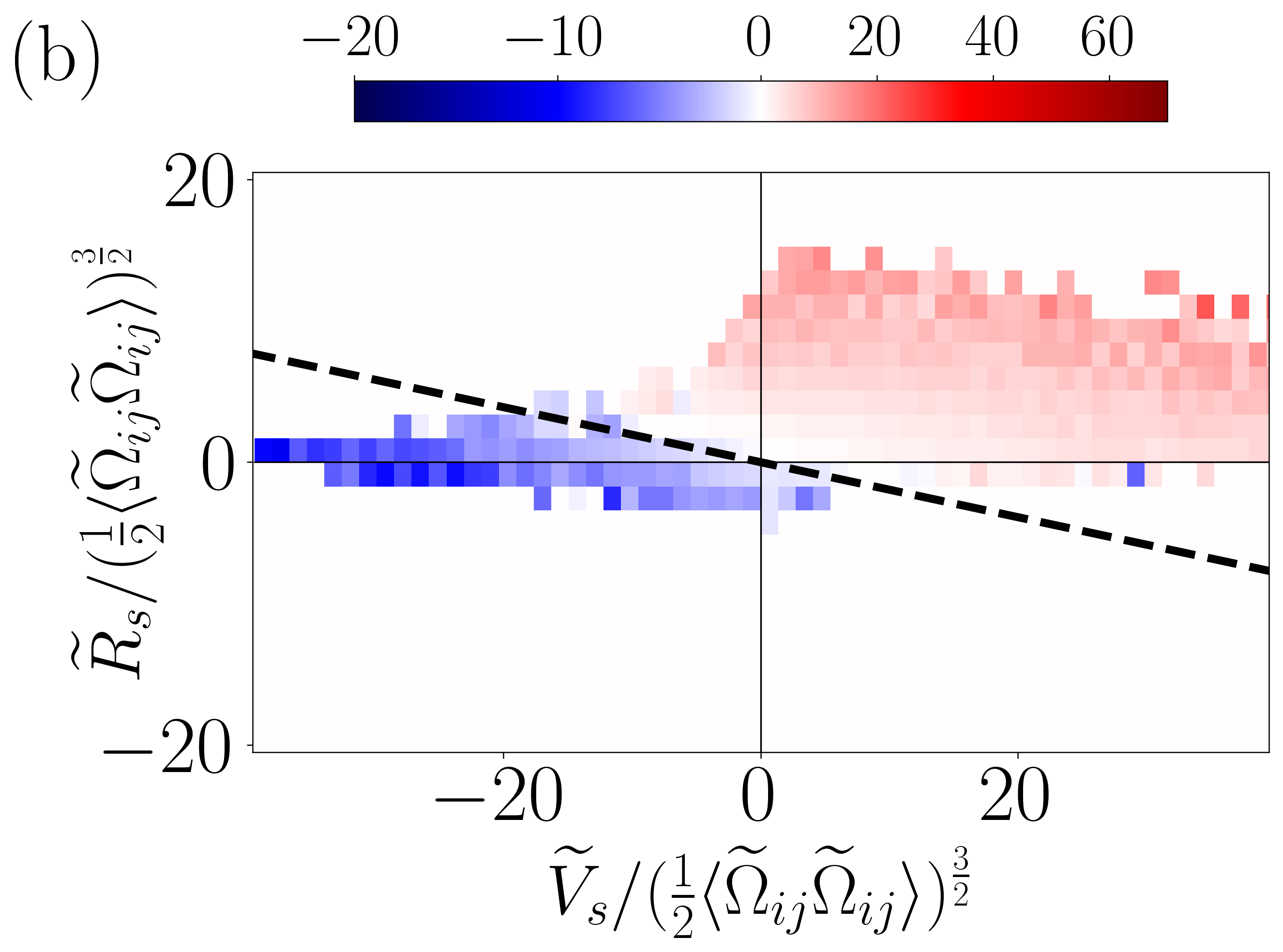}
    \caption{Joint conditional averages at $\ell=15\eta$ of
    (a) the longitudinal energy-transfer rate
    $\langle\Phi_{\ell}^{\parallel}\mid
    \widetilde R_s,\widetilde V_s\rangle$ and
    (b) the transverse energy-transfer rate
    $\langle\Phi_{\ell}^{\perp}\mid
    \widetilde R_s,\widetilde V_s\rangle$.
    Red and blue regions denote positive (forward) and negative (inverse)
    conditional mean transfer, respectively. The black dashed lines represent the analytical prediction of forward and inverse cascade boundary of the longitudinal and transverse energy fluxes:
    $\widetilde R_s=0$ in (a) and
    $\widetilde V_s/10+18\widetilde R_s/35=0$ in (b).
    The normalization is defined in the text.}
    \label{LTOD_15}
\end{figure}

\begin{figure}[htbp]
    \centering
    \includegraphics[width=0.235\textwidth]{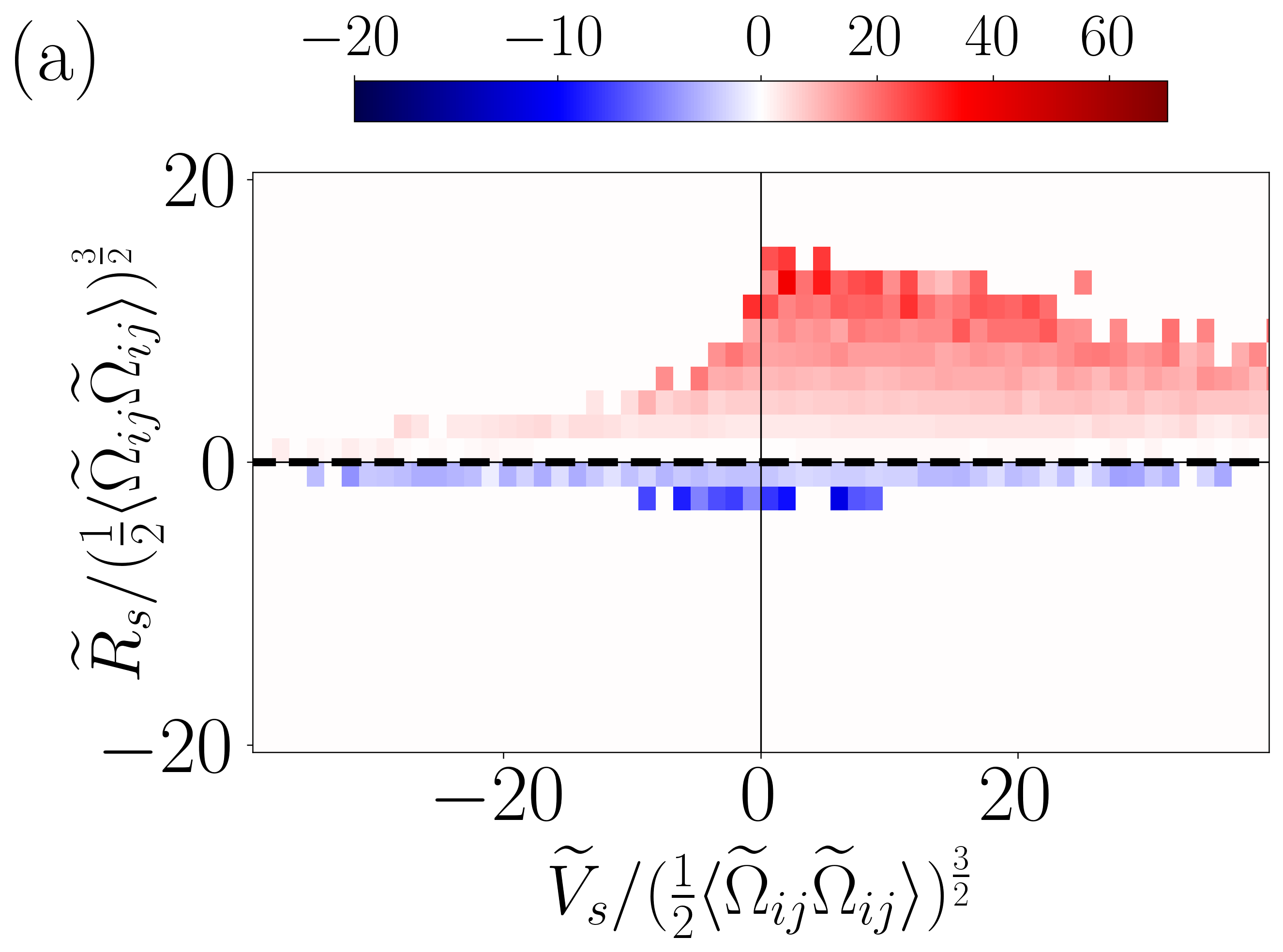}
    \hfill
    \includegraphics[width=0.235\textwidth]{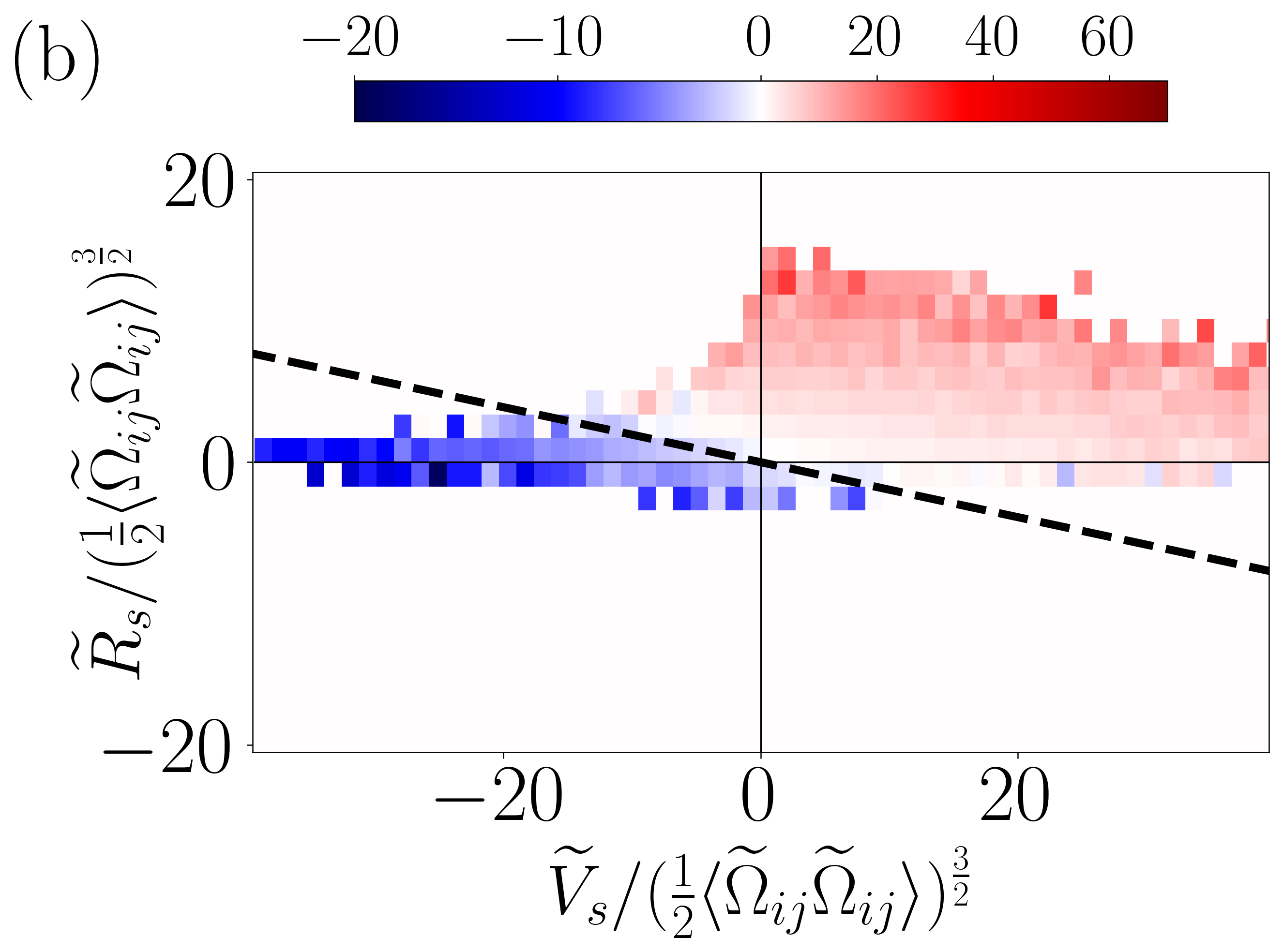}
    \caption{Same as Fig.~\ref{LTOD_15}, but for
    $\ell=30\eta$.}
    \label{LTOD_30}
\end{figure}

\begin{figure}[htbp]
    \centering
    \includegraphics[width=0.235\textwidth]{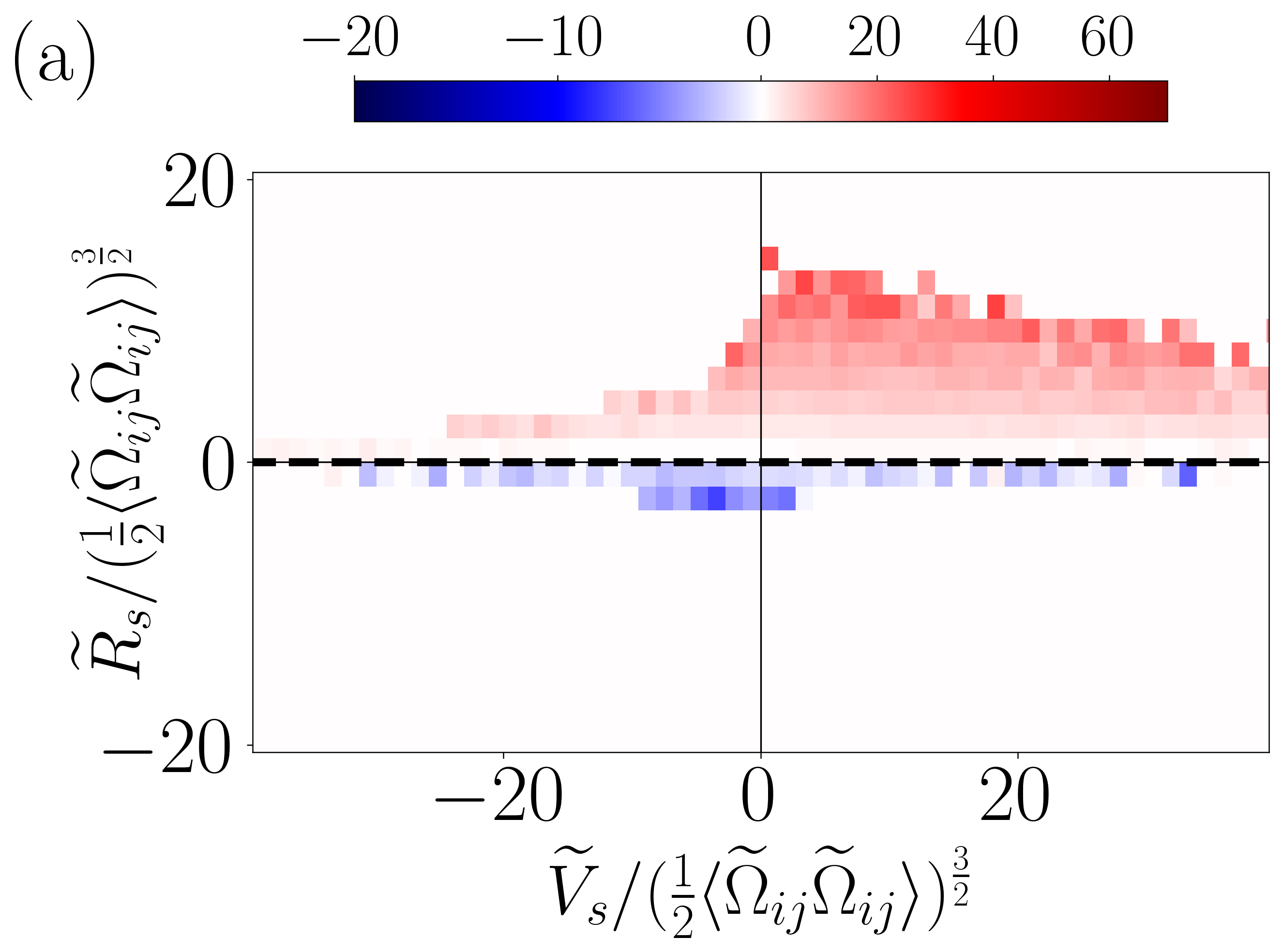}
    \hfill
    \includegraphics[width=0.235\textwidth]{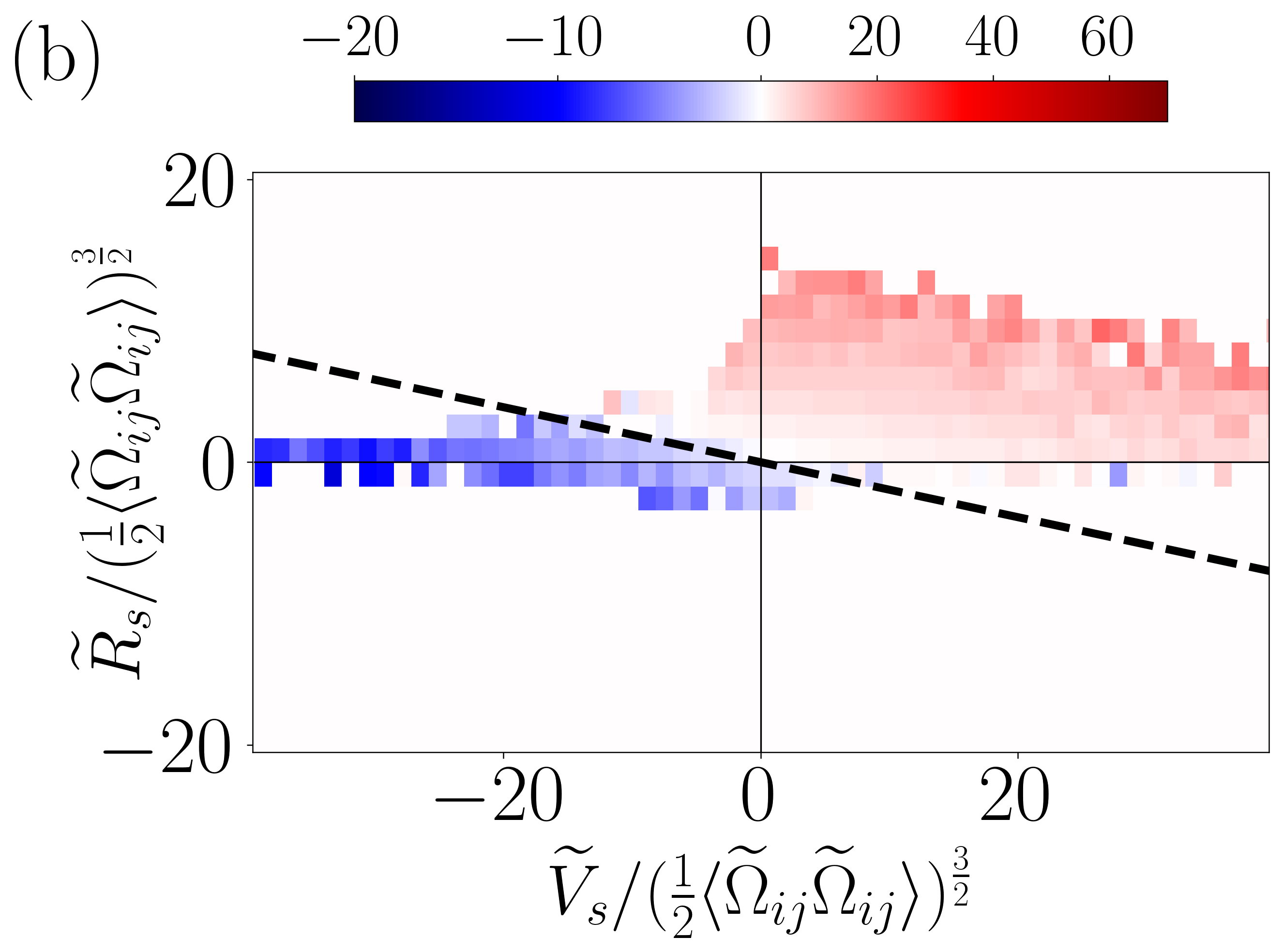}
    \caption{Same as Fig.~\ref{LTOD_15}, but for
    $\ell=60\eta$.}
    \label{LTOD_60}
\end{figure}

\begin{figure}[htbp]
    \centering
    \includegraphics[width=0.235\textwidth]{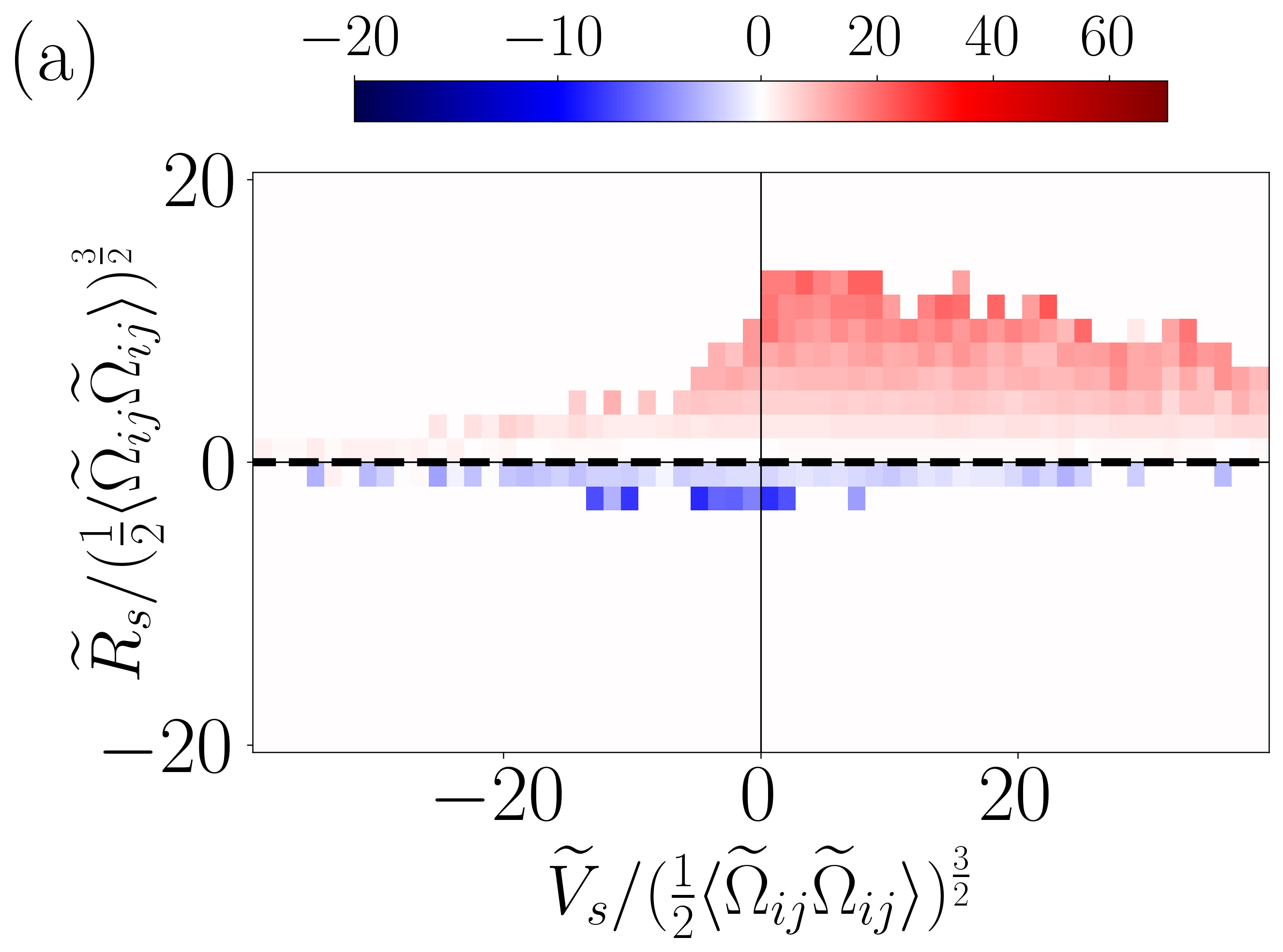}
    \hfill
    \includegraphics[width=0.235\textwidth]{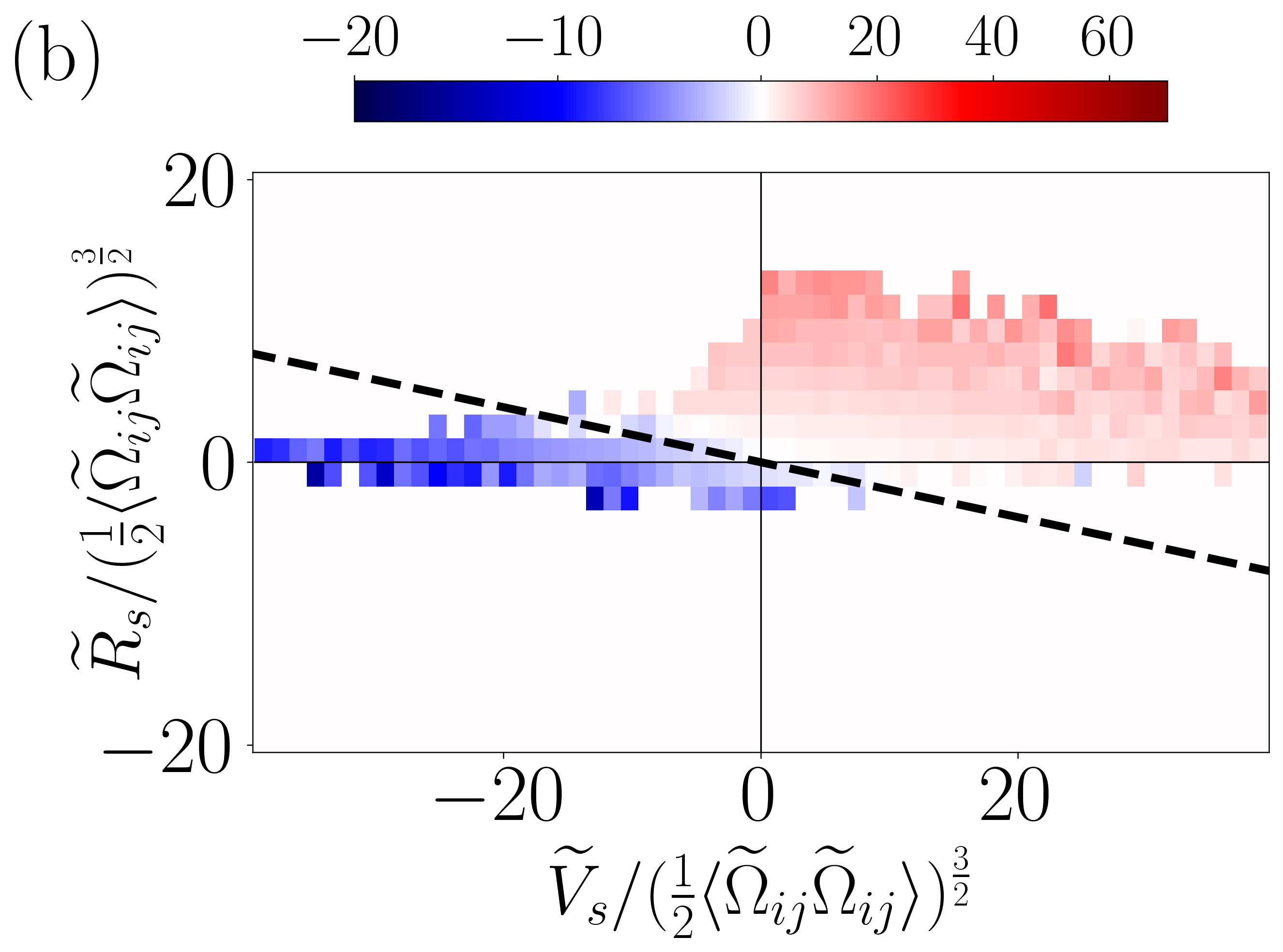}
    \caption{Same as Fig.~\ref{LTOD_15}, but for
    $\ell=75\eta$.}
    \label{LTOD_75}
\end{figure}

Figures~\ref{Con_stat_15}--\ref{Con_stat_75} show the joint
conditional averages
$\langle\Phi_\ell\mid\widetilde R_s,\widetilde V_s\rangle$
at the four additional scales.  Figures~\ref{LTOD_15}--\ref{LTOD_75} show
$\langle\Phi_\ell^{\parallel}\mid\widetilde R_s,\widetilde V_s\rangle$
and
$\langle\Phi_\ell^{\perp}\mid\widetilde R_s,\widetilde V_s\rangle$
at the same four scales. 

\clearpage

\bibliographystyle{apsrev4-2}
\bibliography{2D_PRL_2026}